\documentclass[
 reprint,
 superscriptaddress,
 amsmath,amssymb,
 aps,
]{revtex4-2}

\usepackage{graphicx}% Include figure files
\usepackage{dcolumn}% Align table columns on decimal point
\usepackage{bm}% bold math
\usepackage[colorlinks=true, citecolor=Blue, linkcolor=Navy, urlcolor=RoyalBlue]{hyperref}% add hypertext capabilities
\usepackage{multirow} 
\newcolumntype{P}[1]{>{\centering\arraybackslash}p{#1}}
\newcolumntype{M}[1]{>{\centering\arraybackslash}m{#1}}

\usepackage{diagbox}
\usepackage[svgnames,table]{xcolor}

\usepackage{caption}
\usepackage{subcaption}
\usepackage{ragged2e}

\usepackage{braket}
\usepackage{diagbox}  
\usepackage{makecell}

\DeclareCaptionJustification{justified}{\justifying}

\newcommand{\HG}{\mathrm{HG}}
\newcommand{\LG}{\mathrm{LG}}
\newcommand{\LGmode}[2]{\ensuremath{\LG_{#1,#2}}}

\newcommand{\HGmode}[2]{\ensuremath{\HG_{#1,#2}}}

\begin{document}

%\preprint{APS/123-QED}

\title{Sensitivity of Laguerre-Gaussian Modes to Misalignment and Mode Mismatch in Gravitational-Wave Detectors}

\author{Liu Tao}
\email{liu.tao@apc.in2p3.fr}
\affiliation{Universit\'e Paris Cit\'e, CNRS, Astroparticule et Cosmologie, F-75013 Paris, France}

\author{Matteo Barsuglia}
\affiliation{Universit\'e Paris Cit\'e, CNRS, Astroparticule et Cosmologie, F-75013 Paris, France}

\date{\today}

\begin{abstract}
Higher-order Laguerre-Gaussian (LG) modes form a versatile family of structured optical fields with broader and more uniform transverse intensity distributions than the commonly used fundamental Gaussian mode. These properties make them attractive for precision optical applications, including gravitational-wave detectors, where enhanced spatial averaging of thermally driven test-mass fluctuations can reduce thermal noise. Their practical implementation in precision optical applications, however, requires efficient coupling of the injected beam to the target spatial mode, such as an optical-cavity eigenmode. Residual misalignment and mode mismatch couple power out of the desired spatial mode, thereby reducing the intracavity power buildup and degrading the detector sensitivity. In this work, we analytically and numerically evaluate the power coupling loss induced by misalignment and mode mismatch for a generic $\LGmode{p}{\ell}$ beam. We show that the leading-order loss due to angular or lateral misalignment scales as $2p+|\ell|+1$, while the loss due to waist size or waist position mismatch scales as $2p^2+2p+(2p+1)|\ell|+1$. These results provide a quantitative framework for assessing the coupling robustness of higher-order LG modes in realistic optical cavities. While the sensitivity to imperfect mode coupling generally increases with mode order, the donut-shaped $\LGmode{0}{\ell}$ family exhibits a particularly favorable mode-mismatch scaling, with the loss factor reducing to $|\ell|+1$ and therefore increasing only linearly with the azimuthal index. This enhanced robustness to mode mismatch, together with their broader intensity profiles and large central dark regions that enable selective mirror masking, as suggested in recent work, provides an additional practical motivation for using $\LGmode{0}{\ell}$ modes in precision interferometers such as gravitational-wave detectors.

\end{abstract}

\maketitle

%\tableofcontents

\section{Introduction}
Test-mass thermal noise, originating from thermally driven fluctuations in both the mirror coatings and bulk substrates, poses a major limitation to the sensitivity of current and next-generation gravitational-wave detectors~\cite{Capote_2025, Acernese_2015, CEHorizonStudy, ETDesignReportUpdate2020}. Around the most sensitive frequency band of ground-based interferometers, near $100~\mathrm{Hz}$, it constitutes one of the dominant contributions to the detector noise budget. More broadly, thermal noise also limits the stability and precision of other laser-based experiments employing high-finesse optical cavities~\cite{PhysRevLett.93.250602}, with important implications for optical atomic clocks~\cite{Oelker:2019kqe}, atom interferometry~\cite{DovaleAlvarez:2019ugw}, and searches for dark matter~\cite{Savalle_2021}.

A complementary approach to reducing test-mass thermal noise is to replace the fundamental Gaussian beam with spatial modes having a more uniform intensity distribution~\cite{Vinet2009zz}. By sampling a larger effective area of the mirror surface and substrate, such beams more effectively average over thermally driven displacement fluctuations, as described by the fluctuation-dissipation theorem~\cite{PhysRevD.57.659, Mours_2006, PhysRevD.82.042003}. This approach can reduce the coating and substrate thermal noise contributions without requiring direct changes to the test-mass material properties or mechanical design, in contrast to other thermal noise mitigation strategies such as the development of low mechanical loss coating materials or the operation of silicon test masses under cryogenic conditions~\cite{PhysRevLett.127.071101, Adhikari_2020}.

A well-studied class of spatial beams that is broadly compatible with the existing instrumental infrastructure of current gravitational-wave detectors, including spherical cavity mirror geometries, is formed by higher-order Hermite-Gaussian (HG) and Laguerre-Gaussian (LG) modes. Among these, higher-order LG modes are particularly attractive because their cylindrical symmetry and more azimuthally uniform intensity distributions can provide stronger spatial averaging of thermally driven mirror-displacement fluctuations, and hence greater thermal-noise reduction, than HG modes of the same transverse order~\cite{PhysRevD.82.042003}. In addition, Laguerre-Gaussian modes provide a natural basis for paraxial optical fields with azimuthal symmetry, making them particularly convenient for describing the eigenmodes of optical cavities with spherical mirrors and approximate cylindrical symmetry~\cite{Bond2017}. 

An $\LGmode{p}{\ell}$ mode is specified by the radial index $p \geq 0$, which determines the number of radial nodes, and the azimuthal index $\ell$, which sets the helical phase dependence and gives the number of $2\pi$ phase windings around the beam axis. The corresponding field distribution can be written as
\begin{equation}
\begin{aligned}
\psi_{p,\ell}(r,\phi,z) =\;& 
\frac{1}{w(z)}
\sqrt{\frac{2 p!}{\pi (p+|\ell|)!}}
\left( \frac{\sqrt{2}\, r}{w(z)} \right)^{|\ell|}
L_p^{|\ell|}\!\left( \frac{2 r^2}{w^2(z)} \right) \\
& \times
\exp\!\left[-\frac{r^2}{w^2(z)}
- i \frac{k r^2}{2 R_c(z)}
+ i \ell \phi \right] \\
& \times
\exp\!\left[i(2p+|\ell|+1)\Psi(z)\right],
\end{aligned}
\end{equation}
where $k$ is the wave number, $\Psi(z)$ is the Gouy phase, $L_p^{|\ell|}(x)$ is the generalized Laguerre polynomial, and $w(z)$ and $R_c(z)$ denote the beam radius and wavefront radius of curvature at longitudinal position $z$, respectively.

\begin{figure}[t]
    \centering
    \includegraphics[width=\linewidth]{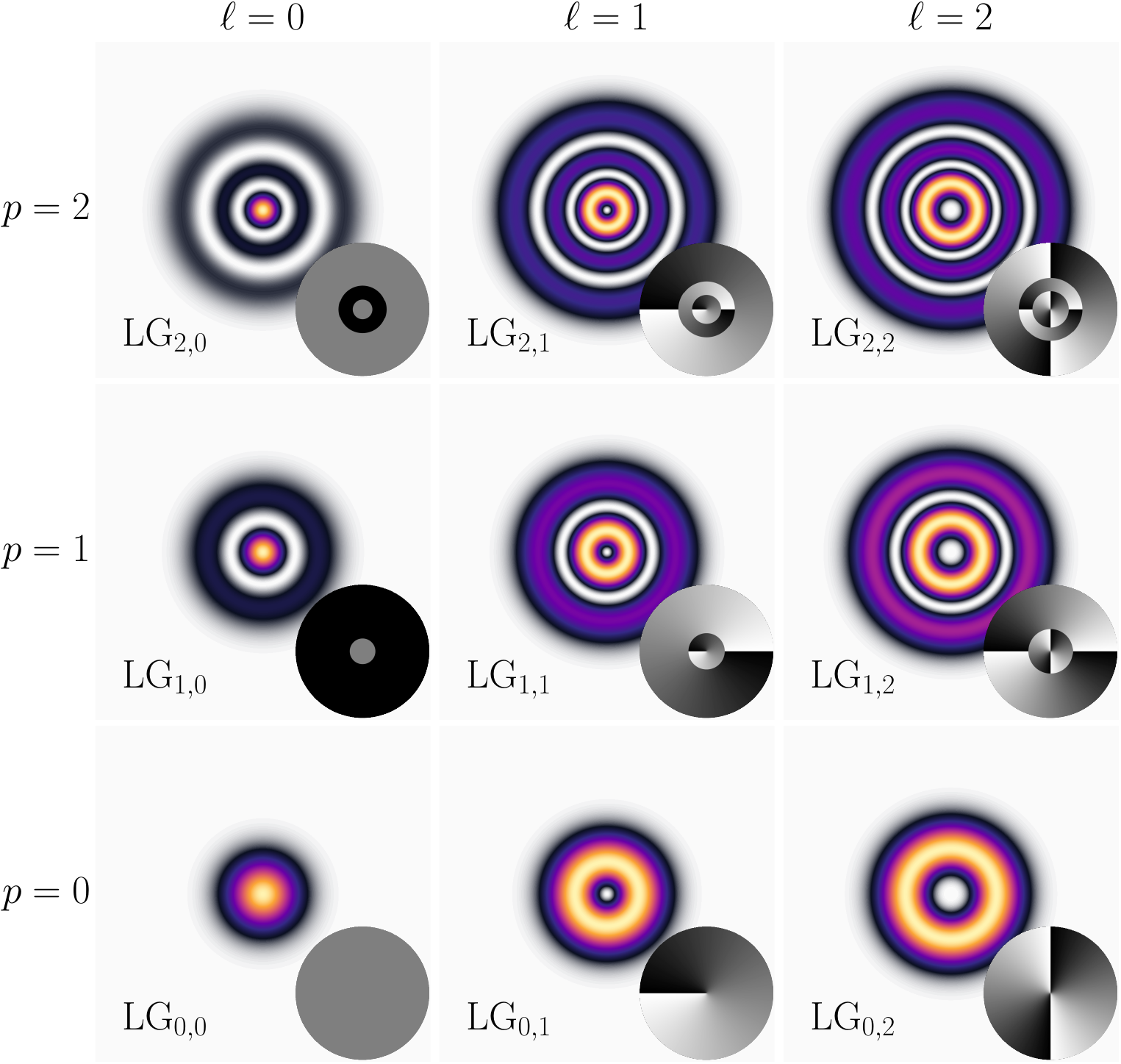}
    \caption{Intensity distributions of LG modes with increasing radial and azimuthal indices $(p,\ell)$. The inset in the lower-right corner of each panel shows the corresponding transverse phase profile, highlighting the azimuthal phase winding and radial phase structure.}
    \label{fig-intensity_phase_inset_LGpl_4x4}
\end{figure}

The transverse order of an LG mode is defined as $\mathcal{N}=2p+|\ell|$. Modes with the same transverse order acquire the same Gouy phase and are therefore frequency degenerate in an ideal optical cavity with spherical mirrors and cylindrical symmetry. The number of modes in a given order is $\mathcal{N}+1 = 2p + |\ell| + 1$, reflecting the number of transverse modes that share the same resonance condition. Fig.~\ref{fig-intensity_phase_inset_LGpl_4x4} shows representative intensity and phase distributions for $\LGmode{p}{\ell}$ modes of increasing order.

These modal properties are particularly relevant to gravitational-wave detectors, where coating thermal noise is a major limiting noise source~\cite{Evans_2008, PhysRevLett127071101}, and the transverse intensity distribution of the optical field determines how thermally driven test-mass fluctuations are sampled and spatially averaged~\cite{PhysRevD.57.659, PhysRevD.82.042003}. By sampling a larger effective mirror area, higher-order LG modes can reduce coating thermal noise. For an $\LGmode{p}{\ell}$ mode, the corresponding coating thermal noise power spectral density reduction factor, relative to the fundamental Gaussian mode, is determined by~\cite{PhysRevD.82.042003}
\begin{equation}
\mathcal{J}_{p,\ell}
=
2 \int_0^{\infty}
e^{-2x}
L_p(x)^2
L_{p+|\ell|}(x)^2
\, dx ,
\end{equation}
where $L_p(x)$ denotes the Laguerre polynomial of order $p$. Tab.~\ref{tab-calculated_matrix} shows the coating thermal noise power spectral density reduction factors for $\LGmode{p}{\ell}$ modes, with the beam size of each mode rescaled to maintain the same 1~ppm clipping loss~\cite{PhysRevD.82.042003}. For example, the coating thermal noise of the sixth-order $\LGmode{2}{2}$ mode is reduced by a factor of 2.373 relative to the fundamental Gaussian mode.

Beyond their potential application in gravitational-wave detectors, higher-order Laguerre-Gaussian modes have also been widely studied in other areas of optics. In particular, modes with nonzero azimuthal index carry orbital angular momentum (OAM), providing an additional spatial degree of freedom for light. A helically phased $\LGmode{p}{\ell}$ beam, with azimuthal phase dependence $\exp(i\ell\phi)$, carries an OAM of $\ell\hbar$ per photon, where $\ell$ is the azimuthal index, or topological charge, $\phi$ is the azimuthal angle, and $\hbar=h/2\pi$ is the reduced Planck constant~\cite{PhysRevA.45.8185}. This property has enabled a broad range of applications, including high-dimensional encoding and entanglement in quantum optics~\cite{Fontaine_2019}, controlled torque and rotation in optical tweezers and micromechanical manipulation~\cite{Porfirev}, and OAM-based multiplexing in optical communications, where the orthogonality of LG modes can be used to increase spectral efficiency and channel capacity~\cite{Wang:2012nkp}.

\begin{table}[t]
    \centering
    \caption{Coating thermal noise power spectral density reduction factors for $\LGmode{p}{\ell}$ modes, with the beam sizes rescaled to maintain the same 1~ppm clipping loss.}
    \begin{tabular}{c|ccc}
        \hline
        \hline
        $p \backslash |\ell| $ & 0 & 1 & 2 \\
        \hline
        0 & 1.000 & 1.396 & 1.595 \\
        1 & 1.655 & 1.976 & 2.133 \\
        2 & 1.925 & 2.230 & 2.373 \\
        \hline
        \hline
    \end{tabular}
    \label{tab-calculated_matrix}
\end{table}

The practical implementation of LG modes in precision interferometers requires that they can be efficiently generated, controlled, and maintained as resonant eigenmodes of optical cavities~\cite{PhysRevD.79.122002}. In precision optical experiments, including laser interferometric gravitational-wave detectors, optical cavities are used both as stable frequency references and as resonant amplifiers of the light-matter interaction~\cite{Kwee_12, Aspelmeyer_2014}. Efficient coupling of the injected laser field to the target cavity eigenmode therefore requires precise control of both alignment and mode matching. Deviations between the input beam axis and the cavity optical axis, as well as mismatches between the beam parameters of the injected field and those of the cavity eigenmode, couple optical power out of the desired resonant mode and into other unwanted spatial modes~\cite{Bayer-Helms:84}. The resulting coupling loss reduces the available intracavity power buildup and can directly degrade the detector sensitivity. In addition, in interferometers employing squeezed vacuum injection, such losses are particularly important because they also reduce the observable quantum noise suppression~\cite{PhysRevD.93.082004, Toyra_2017, McCuller_2021}. Higher-order modes, in particular Hermite-Gaussian modes, have been shown to exhibit increased sensitivity to misalignment and mode mismatch due to the higher spatial-frequency content of their transverse field distributions~\cite{Jones:20}. Specifically, for an $\HGmode{n}{m}$ mode, the leading-order misalignment-induced power coupling loss scales as $n+m+1$, while the corresponding mode mismatch loss scales as $(n^2+n+m^2+m+2)/2$~\cite{Tao_21_loss}. Quantifying the sensitivity of higher-order Laguerre-Gaussian modes to misalignment and mode mismatch is therefore essential for assessing their practical viability and for defining the alignment and mode matching tolerances required to limit power coupling losses.

In this work, we analytically and numerically evaluate the power coupling loss induced by misalignment and mode mismatch for a generic injected $\LGmode{p}{\ell}$ beam. In the analytical approach, the perturbed field is expanded to first order in the relevant misalignment and mode mismatch parameters, allowing the leading-order complex amplitudes of the neighboring scattered modes and the resulting total coupling loss to be expressed in closed form. We find that higher-order LG modes are generally more sensitive to imperfect coupling than the fundamental mode~\cite{Sorazu_2013}. In particular, the loss induced by angular or lateral misalignment scales as $2p+|\ell|+1$, whereas the loss induced by waist size or waist position mismatch scales as $2p^2+2p+(2p+1)|\ell|+1$. These power loss scaling relations are independently verified through numerical calculations. In the numerical approach, the unperturbed and perturbed optical fields are represented as complex two-dimensional arrays, and the power overlap is computed numerically. The corresponding power loss factors are then obtained from the second-order numerical derivatives of the power loss with respect to the relevant perturbation parameters.

These results provide a quantitative framework for assessing the practical feasibility of higher-order LG modes in precision interferometers employing optical cavities, where residual alignment and mode-matching errors are unavoidable. They are particularly relevant for recently proposed donut-shaped $\LGmode{0}{\ell}$ beams, which can be combined with tailored mirror coatings, such as a central AR-coated mask, to selectively suppress scattering into degenerate modes of the same order and improve beam quality in the presence of realistic mirror-surface imperfections~\cite{LG06_mask}. For this class of modes, the mode mismatch loss factor reduces to $|\ell|+1$, increasing only linearly with the azimuthal index. This relatively weak scaling makes $\LGmode{0}{\ell}$ modes more robust against mode mismatch than other LG modes of the same order, making them especially favorable for applications that require both reduced thermal noise and controlled mode-mismatch-induced coupling losses, such as next-generation gravitational-wave detectors.

The remainder of this paper is organized as follows. In \S\ref{sec-mis_mm}, we introduce a unified description of the misalignment and mode mismatch degrees of freedom (DoFs) and establish the conventions used throughout the paper. In \S\ref{sec-loss}, we present the analytical and numerical results of the coupling losses induced by misalignment and mode mismatch for generic $\LGmode{p}{\ell}$ modes, with particular attention to the donut-shaped $\LGmode{0}{\ell}$ mode family. Finally, in \S\ref{sec-conclusion}, we summarize the main results and discuss their implications and possible future extensions. Appendix~\S\ref{sec-derivation} provides a detailed step-by-step derivation of the scattered mode amplitudes and coupling loss factors for $\LGmode{p}{\ell}$ modes under all misalignment and mode-mismatch degrees of freedom, obtained by perturbatively expanding the disturbed field to leading order in the corresponding beam imperfection parameter.

\section{Misalignment and mode mismatch \label{sec-mis_mm}}

When an incident laser field is coupled into an optical cavity, imperfect coupling between the input beam and the target cavity eigenmode can arise from both misalignment and mode mismatch. Misalignment refers to a displacement of the input beam axis relative to the cavity optical axis, either through a lateral offset or an angular tilt. Mode mismatch instead describes a discrepancy between the spatial mode parameters of the injected field and those of the cavity eigenmode, most commonly a mismatch in the waist size or waist position. 

\begin{figure}[t]
    \centering
    \includegraphics[width=\linewidth]{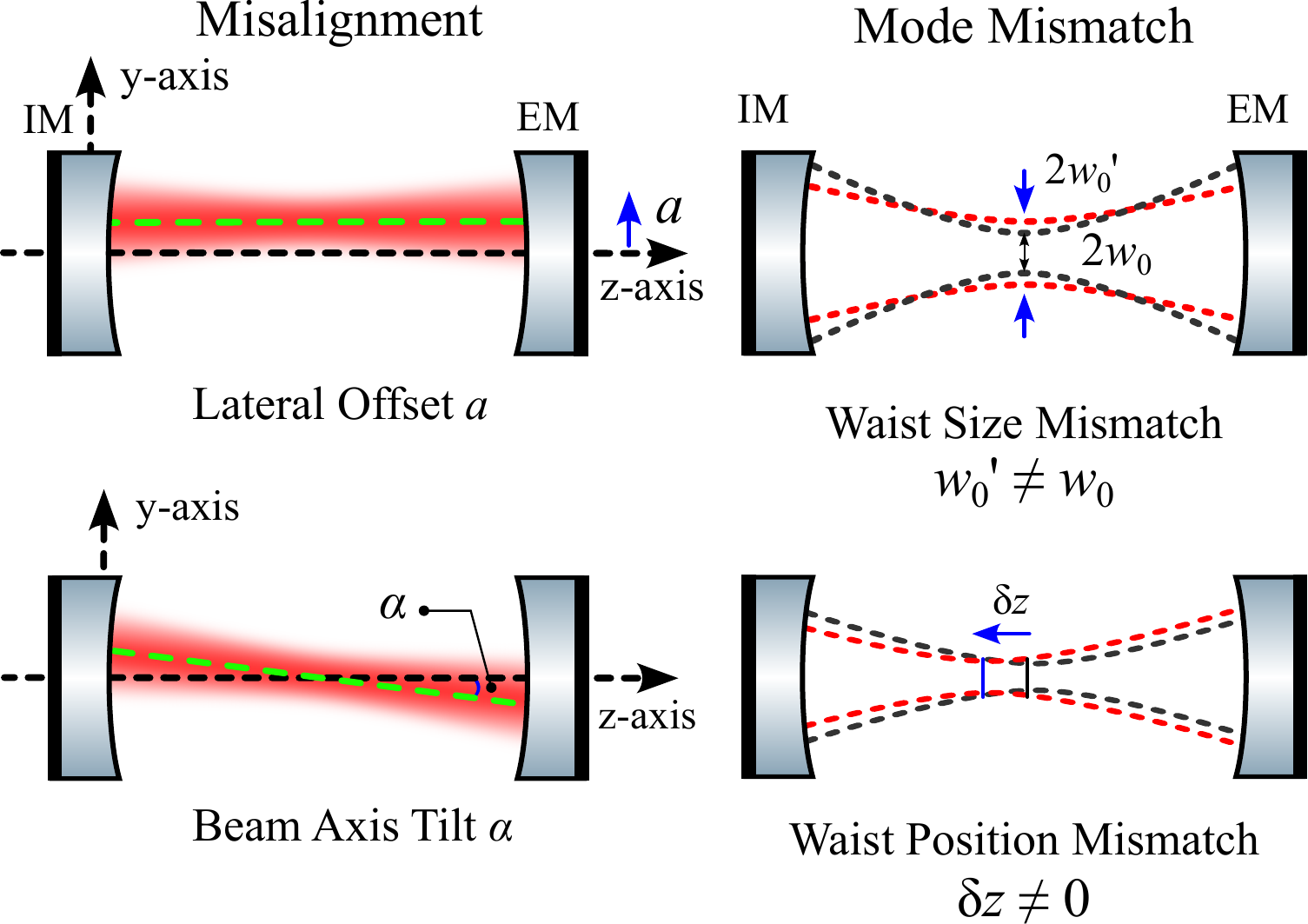}
    \caption{Illustration of beam misalignment (left) and mode mismatch (right) when coupling a laser beam into an optical cavity. Misalignment is described by two orthogonal degrees of freedom: lateral offset $a$ and beam axis tilt $\alpha$. Mode mismatch is similarly described by two orthogonal degrees of freedom: waist size mismatch $\delta w_{0}$ and waist position mismatch $\delta z$.
    }
    \label{fig-illustration_misalignment_MM}
\end{figure}

As illustrated in Fig.~\ref{fig-illustration_misalignment_MM}, these imperfections can be decomposed into a small set of orthogonal geometrical degrees of freedom. A lateral offset corresponds to a transverse displacement $a$ of the beam axis, shown in green, relative to the cavity optical axis, shown in black. An angular misalignment, denoted by $\alpha$, corresponds instead to a tilt of the beam axis with respect to the cavity axis at the waist. Mode mismatch can similarly be described by two independent parameters: a waist size mismatch, for which the injected beam has waist radius $w_0'$ different from the cavity eigenmode waist $w_0$, and a waist position mismatch, characterized by a longitudinal displacement $\delta z$ between the two waist positions. 

\begin{figure*}[t]
    \centering
    \begin{subfigure}{0.49\textwidth}
        \centering
        \includegraphics[width=1\linewidth]{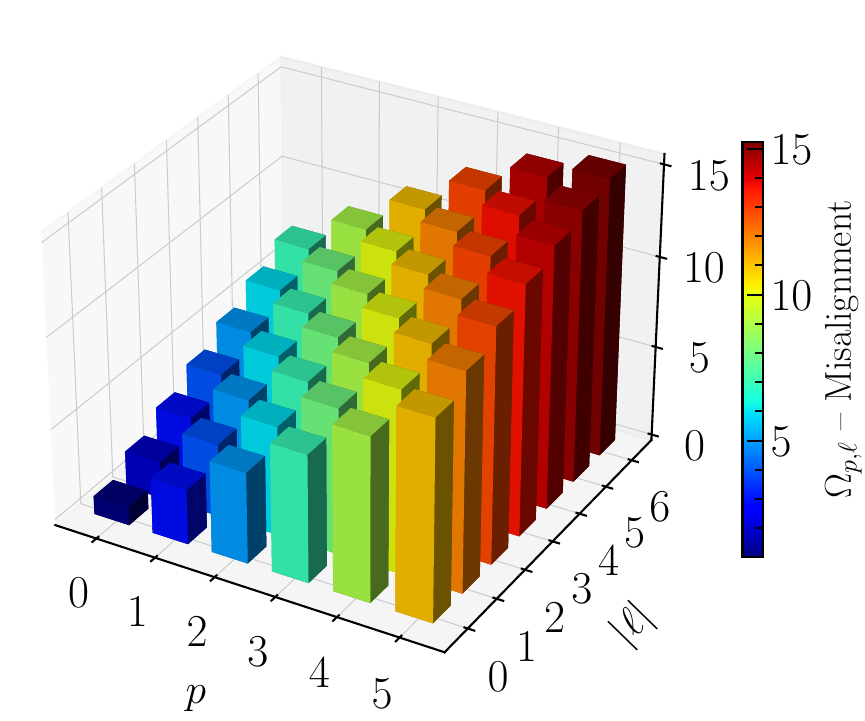}
        \caption{Misalignment (tilt or offset)}
    \end{subfigure}
    \hfill
    \begin{subfigure}{0.49\textwidth}
        \centering
        \includegraphics[width=1\linewidth]{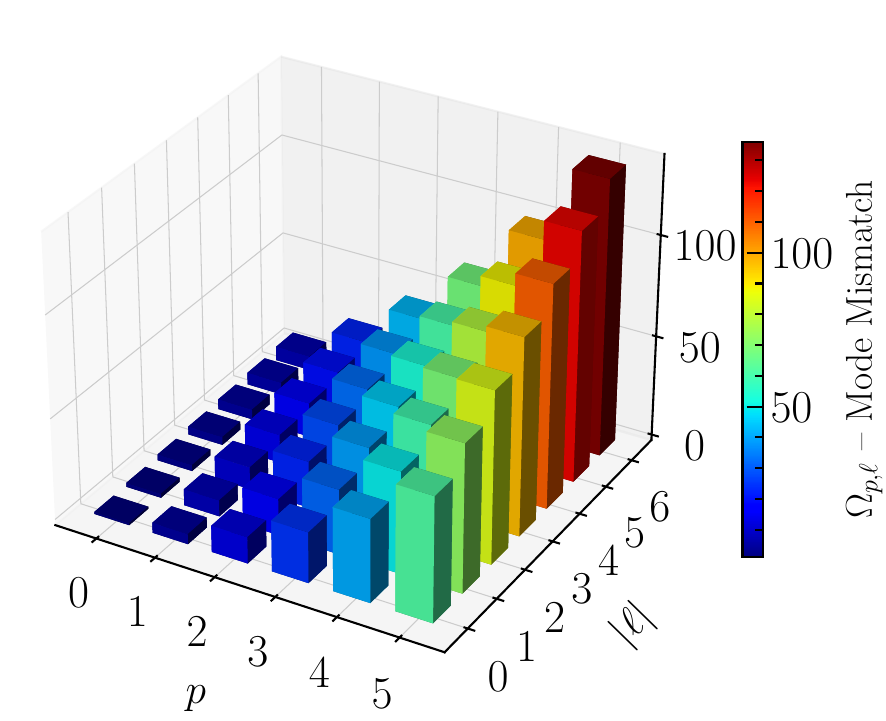}
        \caption{Mode mismatch (waist size or position)}
    \end{subfigure}
    \caption{Power loss factors for $\LGmode{p}{\ell}$ modes induced by imperfect coupling into optical cavities. The left panel shows the loss factor for angular tilt or lateral displacement, while the right panel shows the loss factor for waist position or waist size mismatch. The corresponding analytical scaling relations are summarized in Tab.~\ref{tab-loss_factors}.}
    \label{fig-3d_bar_plot}
\end{figure*}

We introduce dimensionless parameters to describe, in a unified form, the different imperfect coupling degrees of freedom between the input beam and the optical cavity:
\begin{equation}
\begin{array}{rlrl}
\epsilon_{\alpha} &= \dfrac{\alpha}{\Theta}, 
&
\epsilon_{a} &= \dfrac{a}{w_{0}},
\\[6pt]
\epsilon_{z} &= \dfrac{\delta z}{2z_{R}},
&
\epsilon_{w} &= \dfrac{\delta w_{0}}{w_{0}} .
\end{array}
\label{eq-normalized_imperfections}
\end{equation}
The first row corresponds to the two misalignment degrees of freedom, namely angular tilt and lateral displacement, while the second row describes the two mode mismatch degrees of freedom, namely waist position and waist size mismatch. Here, $\Theta=\lambda/(\pi w_0)$ is the far-field divergence angle of the beam, $z_R=\pi w_0^2/\lambda$ is the Rayleigh range, and $\delta w_0 = w_0' - w_0$ denotes the waist size difference between the injected beam and the cavity eigenmode. With these definitions, all imperfect couplings can be characterized in terms of normalized perturbation parameters $\epsilon$.

In the presence of such perturbations, the injected field is no longer perfectly matched to the target cavity eigenmode. Instead, part of the optical power is scattered into other unwanted transverse modes of the cavity basis. This modal coupling leads to power loss from the injected mode, reducing the resonant buildup of the desired cavity eigenmode and, in interferometric applications, degrading the achievable contrast and sensitivity.

\section{Coupling loss scaling for Laguerre-Gaussian modes \label{sec-loss}}

We quantify the power loss associated with imperfect coupling between the injected field and the target cavity eigenmode. Let the unperturbed cavity eigenmode be denoted by $|\psi_{p,\ell}\rangle$, and let the corresponding perturbed input field, including either misalignment or mode mismatch, be denoted by $|\psi_{p,\ell}'\rangle$. The power coupling efficiency is given by the squared modal overlap,
\begin{equation}
\mathcal{P}_{p,\ell} 
=
\left|
\braket{\psi_{p,\ell}|\psi_{p,\ell}'}
\right|^2.
\end{equation}
For a small normalized perturbation parameter $\epsilon$, the perturbed field can be expanded in the orthonormal LG basis as
\begin{equation}
\ket{\psi_{p,\ell}'}
=
a_{p,\ell}\ket{\psi_{p,\ell}}
+
\epsilon \sum_{(p',\ell')\neq(p,\ell)}
c_{p',\ell'}\ket{\psi_{p',\ell'}}
+
\mathcal{O}(\epsilon^2),
\end{equation}
where $a_{p,\ell}$ is the amplitude remaining in the injected mode and $c_{p',\ell'}$ are the first-order amplitudes of the scattered orthogonal modes. Using the orthonormality relation
\begin{equation}
\braket{\psi_{p',\ell'}|\psi_{p'',\ell''}}
=
\delta_{p'p''}\delta_{\ell'\ell''},
\end{equation}
together with the normalization
$\braket{\psi_{p,\ell}'|\psi_{p,\ell}'}=1$, gives
\begin{equation}
1
=
|a_{p,\ell}|^2
+
\epsilon^2
\sum_{(p',\ell')\neq(p,\ell)}
|c_{p',\ell'}|^2
+
\mathcal{O}(\epsilon^3).
\end{equation}
Therefore, the power coupling efficiency is
\begin{equation}
\begin{aligned}
    \mathcal{P}_{p,\ell}
    &=
    \left|
    \braket{\psi_{p,\ell}|\psi_{p,\ell}'}
    \right|^2 = |a_{p,\ell}|^2 
    \\
    &\simeq
    1
    -
    \epsilon^2
    \sum_{(p',\ell')\neq(p,\ell)}
    |c_{p',\ell'}|^2 + \mathcal{O}(\epsilon^3)
    \\
    &=
    1-\epsilon^2\Omega_{p,\ell} + \mathcal{O}(\epsilon^3),
\end{aligned}
\label{eq-overlap}
\end{equation}
where we define
\begin{equation}
    \Omega_{p,\ell}
    =
    \sum_{(p',\ell')\neq(p,\ell)}
    |c_{p',\ell'}|^2
\end{equation}
as the coupling loss factor for the $\LGmode{p}{\ell}$ mode.

The associated power coupling loss is therefore
\begin{equation}
\begin{aligned}
    \mathcal{L}_{p,\ell}
    &=
    1-\mathcal{P}_{p,\ell}
    \simeq
    \epsilon^2\Omega_{p,\ell}
    =
    \epsilon^2
    \sum_{(p',\ell')\neq(p,\ell)}
    |c_{p',\ell'}|^2,
\end{aligned}
\label{eq-powerloss}
\end{equation}
to leading nonvanishing order. Here, $\epsilon$ denotes any of the normalized perturbation parameters introduced in Sec.~\ref{sec-mis_mm}, corresponding to angular tilt, lateral displacement, waist size mismatch, or waist position mismatch. With the normalization chosen in Eq.~\eqref{eq-normalized_imperfections}, it also gives the excess loss relative to the corresponding fundamental mode result for the same normalized beam perturbation. For applications in which the beam sizes of different spatial modes are rescaled to satisfy a common clipping loss requirement, the corresponding mode-dependent beam parameters should be used when evaluating the normalized imperfection parameters in Eq.~\eqref{eq-normalized_imperfections}~\cite{PhysRevD.79.122002}. The coupling loss can then be obtained from Eq.~\eqref{eq-powerloss} using the appropriate beam parameters for each mode. 

\subsection{Analytical derivation}

Appendix~\S\ref{sec-derivation} presents a perturbative derivation of the scattered mode content and the resulting coupling loss factors for $\LGmode{p}{\ell}$ modes under the misalignment and mode mismatch degrees of freedom considered in this work. For each perturbation, the perturbed $\LGmode{p}{\ell}$ field is expanded to first order in the corresponding normalized beam imperfection parameter and decomposed into the LG mode basis, yielding explicit expressions for the complex amplitudes and powers of the scattered modes. By conservation of power, the reduction in the incident mode power equals the total power transferred to the orthogonal scattered modes, as expressed in Eq.~\eqref{eq-powerloss}. Therefore, the leading-order coupling loss is obtained by summing the powers of all modes generated by the first-order perturbative expansion.

For the two misalignment degrees of freedom, angular tilt and lateral displacement, the injected mode $\LGmode{p}{\ell}$ couples to neighboring LG modes with $\ell'=\ell\pm1$ and $\mathcal{N}'=\mathcal{N}\pm1$, where $\ell'$ and $\mathcal{N}'$ denote the azimuthal index and mode order of the scattered mode, respectively. We combine the normalized lateral displacement and angular tilt into the complex misalignment parameter
\begin{equation}
\eta
=
\epsilon_a+i\,\epsilon_\alpha,
\end{equation}
whose real and imaginary parts describe the lateral displacement and angular tilt, respectively. Combining Eqs.~\eqref{eq-LG-expansion-alpha} and~\eqref{eq-LG-expansion-a}, the misalignment-induced scattered mode amplitudes for $\ell>0$ become
\begin{equation}
\begin{array}{rlrl}
c_{p,\ell+1}^{(\eta)}
&=
\dfrac{\eta}{\sqrt{2}}
\sqrt{p+\ell+1},
&
c_{p-1,\ell+1}^{(\eta)}
&=
\dfrac{\eta^{*}}{\sqrt{2}}
\sqrt{p},
\\[6pt]
c_{p,\ell-1}^{(\eta)}
&=
-\dfrac{\eta^{*}}{\sqrt{2}}
\sqrt{p+\ell},
&
c_{p+1,\ell-1}^{(\eta)}
&=
-\dfrac{\eta}{\sqrt{2}}
\sqrt{p+1}.
\end{array}
\end{equation}
These coefficients represent the complex amplitudes of the scattered LG modes. For example, $c_{p,\ell+1}^{(\eta)}$ is the complex amplitude of the scattered mode $\LGmode{p}{\ell+1}$. The corresponding results for $\ell\leq 0$ follow analogously from the relations derived in Appendix~\ref{sec-derivation}.

For simultaneous lateral displacement and angular tilt, the total misalignment-induced coupling loss follows from Eq.~\eqref{eq-powerloss} as
\begin{equation}
\begin{aligned}
\mathcal{L}_{\alpha/a}
&\simeq
\left|c_{p,\ell+1}^{(\eta)}\right|^2
+
\left|c_{p-1,\ell+1}^{(\eta)}\right|^2
+
\left|c_{p,\ell-1}^{(\eta)}\right|^2
+
\left|c_{p+1,\ell-1}^{(\eta)}\right|^2
\\
&=
|\eta|^2
\left(
2p+|\ell|+1
\right)
\\
&=
\left(
\epsilon_a^2+\epsilon_\alpha^2
\right)
\left(
2p+|\ell|+1
\right).
\end{aligned}
\end{equation}
Setting either $\epsilon_a=0$ or $\epsilon_\alpha=0$ recovers the individual tilt- or displacement-induced loss given in Eqs.~\eqref{eq-tilt-all-ell} and~\eqref{eq-lateral-all-ell}. The corresponding misalignment coupling-loss factor is
\begin{equation}
\Omega_{p,\ell}^{\alpha/a}
=
2p+|\ell|+1.
\label{eq-omega_lg_misalignment}
\end{equation}

The two mode mismatch degrees of freedom, waist size and waist position mismatch, preserve the azimuthal index and change the radial index by one, $p'=p\pm1$. The scattered modes are therefore separated from the injected mode by two mode orders, $\mathcal{N}'=\mathcal{N}\pm2$. We combine these degrees of freedom into the complex mode mismatch parameter
\begin{equation}
\mu
=
\epsilon_w+i\,\epsilon_z,
\end{equation}
whose real and imaginary parts describe the waist size and waist position mismatch, respectively. Combining Eqs.~\eqref{eq-LG-expansion-w} and~\eqref{eq-LG-expansion-z}, the scattered-mode amplitudes are
\begin{equation}
\begin{array}{rl}
c_{p+1,\ell}^{(\mu)}
&=
-\mu^{*}
\sqrt{(p+1)(p+|\ell|+1)},
\\[6pt]
c_{p-1,\ell}^{(\mu)}
&=
\mu
\sqrt{p(p+|\ell|)},
\end{array}
\end{equation}
for $\LGmode{p+1}{\ell}$ mode and $\LGmode{p-1}{\ell}$ mode. Because these expressions depend on the azimuthal index only through $|\ell|$, they apply directly to arbitrary $\ell$.

For simultaneous waist size and waist position mismatch, the total mode-mismatch-induced power coupling loss follows from Eq.~\eqref{eq-powerloss} as
\begin{equation}
\begin{aligned}
\mathcal{L}_{w/z}
&\simeq
\left|c_{p+1,\ell}^{(\mu)}\right|^2
+
\left|c_{p-1,\ell}^{(\mu)}\right|^2
\\
&=
|\mu|^2
\left[
2p^2+2p+(2p+1)|\ell|+1
\right]
\\
&=
\left(
\epsilon_w^2+\epsilon_z^2
\right)
\left[
2p^2+2p+(2p+1)|\ell|+1
\right].
\end{aligned}
\end{equation}
Setting either $\epsilon_z=0$ or $\epsilon_w=0$ recovers the individual waist size or waist position mismatch loss given in Eqs.~\eqref{eq-ws-all-ell} and~\eqref{eq-wp-all-ell}. The corresponding mode mismatch coupling loss factor is
\begin{equation}
\Omega_{p,\ell}^{w/z}
=
2p^2+2p+(2p+1)|\ell|+1.
\label{eq-omega_lg_modemismatch}
\end{equation}

\begin{table}[t]
\centering
\caption{Misalignment and mode mismatch induced power loss factors for generic $\LGmode{p}{\ell}$ modes, the donut-shaped $\LGmode{0}{\ell}$ family, and $\HGmode{n}{m}$ modes.}
\begin{tabular}{ccc}
\hline
\hline
Mode & Misalignment & Mode mismatch \\
\hline
$\LGmode{p}{\ell}$ 
& $2p + |\ell| + 1$ 
& $2p^2 + 2p + (2p+1)|\ell| +1$ 
\\
$\LGmode{0}{\ell}$ 
& $|\ell| + 1$ 
& $|\ell| + 1$ 
\\
$\HGmode{n}{m}$ 
& $n + m + 1$ 
& $(n^2+n+m^2+m+2)/2$
\\
\hline
\hline
\end{tabular}
\label{tab-loss_factors}
\end{table}

This loss factor scales quadratically with the radial index $p$, but only linearly with the azimuthal index $|\ell|$. For the special case $p=0$, corresponding to the donut-shaped $\LGmode{0}{\ell}$ family, the mode mismatch loss factor reduces to
\begin{equation}
    \left.\Omega_{p,\ell}^{w/z}\right|_{p=0}
    =
    \Omega_{0,\ell}^{w/z}
    =
    |\ell|+1.
\end{equation}
Thus, for this class of modes, the sensitivity to mode mismatch increases only linearly with the azimuthal index.

The analytical loss factors are summarized in Tab.~\ref{tab-loss_factors} and shown in Fig.~\ref{fig-3d_bar_plot} for modes with $p \leq 5$ and $|\ell| \leq 6$. These scaling factors are consistent with earlier experimental observations for an $\LGmode{3}{3}$ beam coupled to an optical cavity, where the sensitivity to residual misalignment and mode-matching errors was reported to be enhanced by factors of approximately $8$ and $40$, respectively, relative to a fundamental Gaussian beam with the same Gaussian parameters~\cite{PhysRevD.90.122011}. For the same $\LGmode{3}{3}$ mode, the analytical expressions in Eqs.~\eqref{eq-omega_lg_misalignment} and~\eqref{eq-omega_lg_modemismatch} give $\Omega^{\alpha/a}_{3,3}=10$ and $\Omega^{w/z}_{3,3}=46$, in good agreement with these experimental trends.

For comparison, the corresponding power-loss factors due to misalignment and mode mismatch for Hermite-Gaussian $\HGmode{n}{m}$ modes are also included~\cite{Tao_21_loss}. Interestingly, for misalignment, the loss factors for both LG and HG modes scale with the number of co-resonant modes,
\begin{equation}
\mathcal{N}+1 = 2p+|\ell|+1 = n+m+1 .
\end{equation}
For misalignment, LG and HG modes of the same transverse order share the same power loss factor. The donut-shaped \LGmode{0}{\ell} modes therefore have neither an additional advantage nor a disadvantage compared with other modes of the same order.

In contrast, the mode mismatch loss shows a different dependence on the mode indices. For HG modes, the loss factor is given by $(n^2+n+m^2+m+2)/2$, which increases quadratically with each transverse mode index, $n$ and $m$. This indicates that the sensitivity to mode mismatch grows rapidly for higher-order HG modes. For LG modes, however, the quadratic dependence appears \textit{only} in the radial index $p$, while the dependence on the azimuthal index $|\ell|$ remains linear.

This behavior is further illustrated in Fig.~\ref{fig-3d_bar_plot}: the mode mismatch loss factors increase slowly, with a linear dependence along the $|\ell|$ direction, but grow much more rapidly, with a quadratic dependence along the $p$ direction. The weaker dependence on $|\ell|$ is especially important for the donut-shaped $\LGmode{0}{\ell}$ family, for which the mode mismatch loss factor is simply $|\ell|+1$. As a result, among higher-order modes of the same order, the $\LGmode{0}{\ell}$ modes exhibit a more moderate sensitivity to mode mismatch than modes with nonzero radial index. 

\subsection{Numerical validation}

As an independent check of the analytical results in Tab.~\ref{tab-loss_factors}, we also evaluate the coupling loss factors numerically. The power overlap in Eq.~\eqref{eq-overlap} and the corresponding power coupling loss in Eq.~\eqref{eq-powerloss} are computed by representing the target cavity eigenmode $|\psi_{p,\ell}\rangle$ and the perturbed input field $|\psi_{p,\ell}'\rangle$ as two-dimensional complex field arrays. For each misalignment or mode mismatch degree of freedom, the perturbation parameter $\epsilon$ is varied around $\epsilon=0$, and the resulting coupling loss curve is evaluated directly from the numerical field overlap.

\begin{figure}[t]
    \centering
    \includegraphics[width=\linewidth]{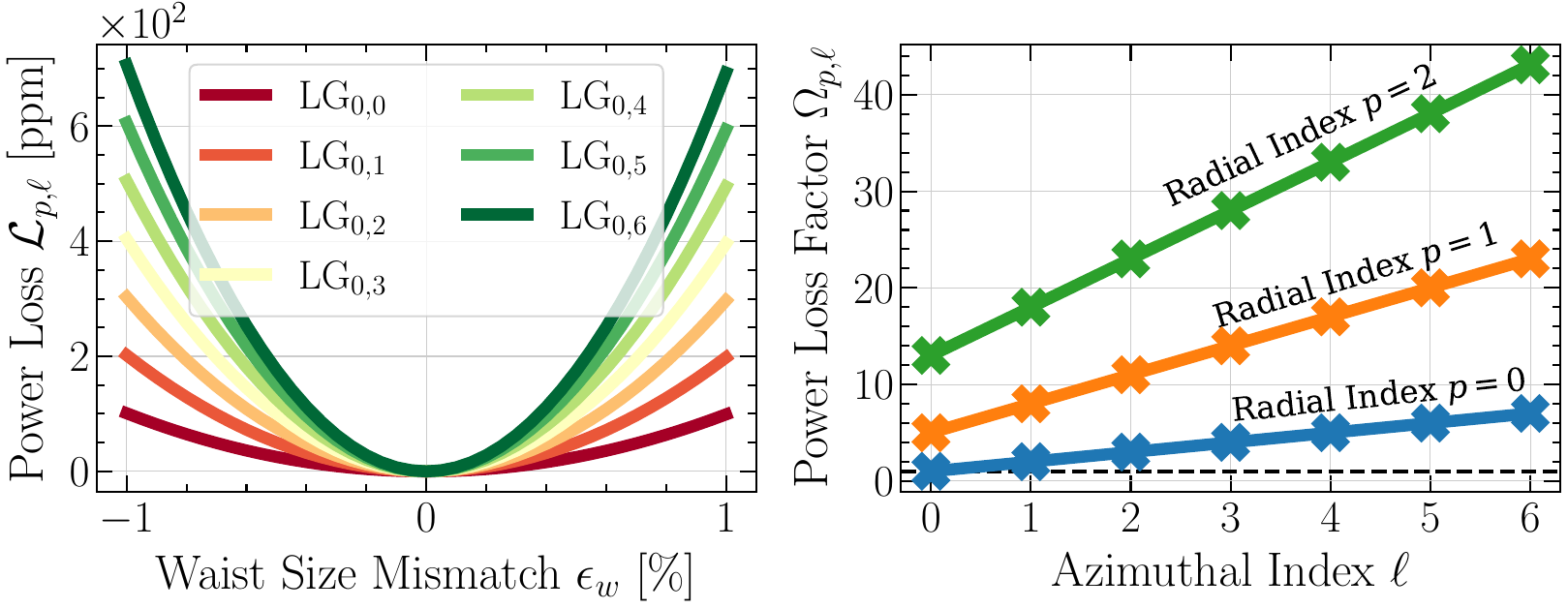}
    \caption{\textit{Left:} Power coupling loss induced by waist size mismatch for $\LGmode{0}{\ell}$ modes with $\ell=0$--6. \textit{Right:} Corresponding mode mismatch loss factors for higher-order LG modes as functions of the azimuthal index $\ell$. The three curves correspond to radial indices $p=0$, 1, and 2.}
    \label{fig-power_loss_mismatch_size}
\end{figure}

The loss factor $\Omega_{p,\ell}$ can then be extracted from the curvature of the coupling loss curve at $\epsilon=0$,
\begin{equation}
\Omega_{p,\ell}
=
\frac{1}{2}
\left.
\frac{\partial^2 \mathcal{L}_{p,\ell}}{\partial \epsilon^2}
\right|_{\epsilon=0}.
\label{eq-factor_numerical}
\end{equation}
This procedure provides a direct numerical extraction of the leading-order power loss coefficient in Eq.~\eqref{eq-powerloss}, without relying on the analytical perturbative expansion. It also provides a simple and more general numerical framework for evaluating power coupling losses due to generic beam imperfections, especially in cases where the analytical perturbative treatment developed in this work becomes cumbersome.

\begin{figure}[t]
    \centering
    \includegraphics[width=\linewidth]{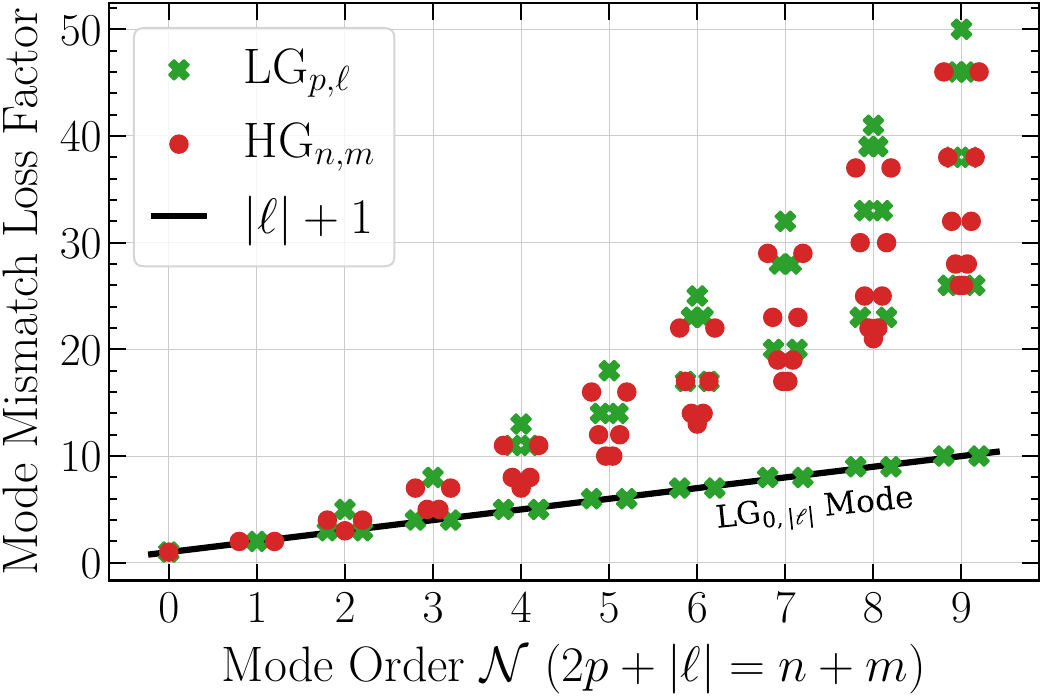}
    \caption{Mode mismatch-induced power loss factors for HG and LG modes over a range of transverse mode order $\mathcal{N}$. The donut-shaped $\LGmode{0}{\ell}$ modes follow the simplified scaling relation $|\ell|+1$ and exhibit the lowest power loss factors among higher-order modes with the same mode order, as indicated by the black line.
    }
    \label{fig-mode_mismatch_loss_LG_HG}
\end{figure}

As an example, the left panel of Fig.~\ref{fig-power_loss_mismatch_size} shows the waist size mismatch-induced power coupling loss, expressed in ppm, for $\LGmode{0}{\ell}$ modes with $\ell=0$--6. For a fixed value of the normalized waist size mismatch $\epsilon_w$, the coupling loss increases with the azimuthal index, illustrating the enhanced sensitivity of higher-order modes to imperfect mode matching coupling. The right panel shows the corresponding loss factors extracted using Eq.~\eqref{eq-factor_numerical} for LG modes with different radial and azimuthal indices. The numerical results reproduce the linear dependence on $|\ell|$ for fixed $p$, as well as the stronger dependence on the radial index predicted in Tab.~\ref{tab-loss_factors}. The same numerical procedure was applied to all four perturbation degrees of freedom, and the extracted loss factors agree with the analytical scaling relations summarized in Tab.~\ref{tab-loss_factors}.

Fig.~\ref{fig-mode_mismatch_loss_LG_HG} compares the mode mismatch-induced power loss factors for $\LGmode{p}{\ell}$ modes, shown in green crosses, and $\HGmode{n}{m}$ modes, shown in red dots, over a range of transverse mode orders $\mathcal{N}$. The spread of mode-mismatch loss factors for HG modes of the same mode order is smaller than that for LG modes. In particular, at fixed transverse order $\mathcal{N}=n+m$, the HG mode-mismatch loss factor is minimized when the mode indices are as nearly equal as possible: $n=m$ for even $\mathcal{N}$ and $|n-m|=1$ for odd $\mathcal{N}$. Consequently, the symmetric $\HGmode{n}{n}$ family, such as the well-studied $\HGmode{3}{3}$ mode, which is attractive for gravitational-wave detectors because of its favorable spatial symmetry and thermal-noise performance, is the least susceptible to mode mismatch among HG modes of the same even transverse order.

In contrast, the LG mode family contains modes with either particularly large or particularly small mode mismatch loss factors. The lowest group of loss factors corresponds to the donut-shaped $\LGmode{0}{\ell}$ family, which follows the linear scaling $|\ell|+1$, indicated by the black line. These modes therefore exhibit the greatest robustness to mode mismatch, with loss factors substantially smaller than those of \textit{all} other LG and HG modes of the same transverse order. For example, at sixth order, the loss factors for the commonly studied $\LGmode{2}{2}$ and $\HGmode{3}{3}$ modes are 23 and 13, respectively, whereas the corresponding factor for $\LGmode{0}{6}$ is only 7. Similarly, at the ninth order, the loss factors for $\LGmode{3}{3}$ and $\HGmode{4}{5}$ are 46 and 26, respectively, compared with only 10 for $\LGmode{0}{9}$. This reduced sensitivity to mode mismatch provides a practical advantage for the $\LGmode{0}{\ell}$ family in optical cavity applications where residual mode matching errors are unavoidable.

\begin{table}[t]
\centering
\caption{Parameters of symmetric $3\,\mathrm{km}$ cavities designed to give a clipping loss of $1\,\mathrm{ppm}$ for the $\LGmode{0}{0}$ and $\LGmode{0}{6}$ modes. Here, $w_{\mathrm{m}}$ and $w_0$ denote the Gaussian beam size at the mirrors and the beam waist size, respectively.}
\label{tab-equal_clipping_cavities}
\begin{ruledtabular}
\begin{tabular}{lcc}
Parameter & $\LGmode{0}{0}$ & $\LGmode{0}{6}$ \\
\hline
$w_{\mathrm{m}}$ [cm] & $6.47$ & $4.60$ \\
$w_0$ [cm]            & $0.791$ & $1.14$ \\
$z_R$ [m]              & $184.7$ & $383.1$ \\
$R$ [m]                & $1521.8$ & $1596.9$ \\
$g_1g_2$               & $0.94$ & $0.77$ \\
\end{tabular}
\end{ruledtabular}
\end{table}

\subsection{Loss Scaling at Equal Clipping}

The intrinsic power loss factors obtained above compare different spatial modes under the assumption that they share the same Gaussian beam parameter and, consequently, the same normalized optical imperfections defined in Eq.~\eqref{eq-normalized_imperfections}. In applications aimed at reducing test-mass thermal noise, such as gravitational-wave detectors, however, the beam parameters of different spatial modes are typically selected to satisfy a fixed clipping-loss requirement~\cite{PhysRevD.79.122002}. Because higher-order modes have intensity distributions that extend farther toward the outer region of the test mass, satisfying the same clipping-loss constraint requires a smaller Gaussian-envelope beam size at the cavity mirrors. The associated changes in the cavity geometry, waist size, and Rayleigh range modify the normalized imperfection parameters in Eq.~\eqref{eq-normalized_imperfections} and, consequently, the relative power losses. The intrinsic power loss factors derived in Eqs.~\eqref{eq-omega_lg_misalignment} and~\eqref{eq-omega_lg_modemismatch} must therefore be rescaled using the mode-dependent beam parameters.

As an illustrative example, we consider a symmetric Virgo-like arm cavity with a length of $L=3\,\mathrm{km}$, a test mass radius of $R_{\mathrm{TM}}=0.17\,\mathrm{m}$, and a wavelength of $\lambda=1064\,\mathrm{nm}$. Requiring the clipping loss $\mathcal{L}_{\mathrm{clip}}=1\,\mathrm{ppm}$ gives the beam size $w_{\mathrm{m}}=6.47\,\mathrm{cm}$ for the fundamental $\LGmode{0}{0}$ mode and $w_{\mathrm{m}}=4.60\,\mathrm{cm}$ for the higher-order $\LGmode{0}{6}$ mode~\cite{LG06_mask}. The corresponding waist sizes, Rayleigh ranges, mirror radii of curvature, and cavity $g$ factors are calculated using the standard Gaussian beam relations for a symmetric near-concentric cavity and are summarized in Tab.~\ref{tab-equal_clipping_cavities}.

For the $\LGmode{0}{6}$ mode, the intrinsic misalignment and mode mismatch loss factors relative to the fundamental $\LGmode{0}{0}$ mode are both
\begin{equation}
\Omega_{0,6}^{\alpha/a}
=
\Omega_{0,6}^{w/z}
=7.
\end{equation}
Thus, for equal beam parameters and equal normalized imperfections, the $\LGmode{0}{6}$ mode experiences seven times the power loss of the fundamental mode. Under the equal clipping loss condition, however, the relevant normalized imperfections must be evaluated using the mode-dependent values of $w_0$ and $z_R$ from Tab.~\ref{tab-equal_clipping_cavities}.

For the same physical angular tilt $\alpha$, lateral offset $a$, waist size error $\delta w_0$, and waist position error $\delta z$, the corresponding normalized parameters scale as
\begin{equation}
\begin{aligned}
\epsilon_\alpha &\propto w_0\alpha,
& \epsilon_a &\propto \frac{a}{w_0},
& \epsilon_z &\propto \frac{\delta z}{z_R},
& \epsilon_w &\propto \frac{\delta w_0}{w_0}.
\end{aligned}
\end{equation}

The rescaled relative power loss factors for the $\LGmode{0}{6}$ mode, normalized to the corresponding $\LGmode{0}{0}$ losses, are therefore
\begin{equation}
\begin{aligned}
\Omega'_\alpha
&= 7\cdot \left(\frac{w_{0,06}}{w_{0,00}}\right)^2,
&
\Omega'_a
&= 7\cdot \left(\frac{w_{0,00}}{w_{0,06}}\right)^2,\\
\Omega'_z
&= 7\cdot \left(\frac{z_{R,00}}{z_{R,06}}\right)^2,
&
\Omega'_w
&= 7\cdot \left(\frac{w_{0,00}}{w_{0,06}}\right)^2.
\end{aligned}
\end{equation}

For the cavity parameters listed in Tab.~\ref{tab-equal_clipping_cavities}, the corresponding ratios are $w_{0,06}/w_{0,00}=1.44$ and $z_{R,06}/z_{R,00}=2.07$. The resulting rescaled power loss factors are summarized in Tab.~\ref{tab-equal_clipping_loss_factors}, assuming that the absolute physical beam perturbations are held fixed between the two cavity configurations.

\begin{table}[t]
\centering
\caption{Power loss of the equal clipping loss $\LGmode{0}{6}$ configuration relative to the $\LGmode{0}{0}$ configuration for the same absolute physical imperfection.}
\label{tab-equal_clipping_loss_factors}
\begin{ruledtabular}
\begin{tabular}{lcc}
Imperfection & Relative factor & Value \\
\hline
Angular tilt $\alpha$
& $7\cdot (w_{0,06}/w_{0,00})^2$
& $14.5$ \\
Lateral offset $a$
& $7\cdot (w_{0,00}/w_{0,06})^2$
& $3.4$ \\
Waist size error $\delta w_0$
& $7\cdot (w_{0,00}/w_{0,06})^2$
& $3.4$ \\
Waist position error $\delta z$
& $7\cdot (z_{R,00}/z_{R,06})^2$
& $1.6$ \\
\end{tabular}
\end{ruledtabular}
\end{table}

The equal clipping loss rescaling therefore produces quantitatively different effects for the four degrees of freedom. The larger waist size of the $\LGmode{0}{6}$ mode increases its normalized angular tilt parameter, raising the tilt-induced loss enhancement from the intrinsic factor of $7$ to approximately $14.5$. In contrast, the same rescaling reduces the normalized lateral displacement and waist size mismatch parameters, lowering their loss enhancements to approximately $3.4$. The larger Rayleigh range also substantially reduces the normalized waist position mismatch, yielding a loss enhancement of approximately $1.6$. Thus, although the $\LGmode{0}{6}$ mode remains more sensitive than $\LGmode{0}{0}$ to all four absolute physical perturbations considered in this example, the intrinsic factor of $7$ alone does not accurately describe its relative sensitivity when the cavity geometry and beam parameters are modified to satisfy the same clipping loss constraint.

\section{Conclusion \label{sec-conclusion}}

Higher-order Laguerre-Gaussian (LG) modes provide a promising route for reducing test-mass thermal noise in precision interferometry, including gravitational-wave detectors, by distributing optical power over a larger effective mirror area. Their practical implementation, however, requires careful control of the coupling between the injected beam and the target cavity eigenmode. Residual misalignment and mode mismatch can scatter power out of the desired resonant mode, reducing the available intracavity power buildup and introducing additional optical loss. Quantifying these effects is therefore essential for assessing the feasibility of higher-order LG modes in realistic optical cavities.

In this work, we analytically and numerically evaluated the power coupling loss induced by misalignment and mode mismatch for a generic $\LGmode{p}{\ell}$ beam. Using a perturbative expansion to first order in the normalized imperfection parameters, we derive closed-form expressions for the scattered mode amplitudes and the corresponding leading-order coupling loss factors. For the two misalignment degrees of freedom, corresponding to angular tilt and lateral displacement, the power loss factor is $2p+|\ell|+1$. For the two mode mismatch degrees of freedom, corresponding to waist position and waist size mismatch, the power loss factor is $2p^2+2p+(2p+1)|\ell|+1$. These results show that, in general, higher-order LG modes are more sensitive to imperfect coupling than the fundamental Gaussian mode.

The analytical scaling relations were independently validated using numerical overlap calculations, in which the unperturbed and perturbed optical fields were represented as two-dimensional complex field arrays. Comparison with Hermite-Gaussian modes further shows that, while the misalignment-induced loss scales with the number of co-resonant modes in both mode bases, the mode mismatch-induced loss exhibits a distinct dependence on the LG mode indices. In particular, whereas the mode mismatch loss factor for $\HGmode{n}{m}$ modes scales quadratically with both transverse indices $n$ and $m$, for $\LGmode{p}{\ell}$ modes the quadratic dependence appears only through the radial index $p$, while the dependence on the azimuthal index $|\ell|$ remains linear.

This behavior is especially relevant for the recently proposed donut-shaped $\LGmode{0}{\ell}$ family. For these modes, the mode mismatch loss factor reduces to $|\ell|+1$, which increases only linearly with the azimuthal index. As a result, $\LGmode{0}{\ell}$ modes exhibit a much more moderate sensitivity to mode mismatch than other LG or HG modes of the same order. This provides an additional practical motivation for considering $\LGmode{0}{\ell}$ modes in optical cavities for gravitational-wave detectors, where their broader spatial profiles can reduce test-mass thermal noise and their donut-shaped intensity distributions may improve robustness against mirror scattering imperfections through tailored mirror coating profiles~\cite{LG06_mask}.

Future work will extend this analysis toward sensing and control strategies for higher-order LG beams, to mitigate the enhanced power coupling losses arising from residual misalignment and mode mismatch~\cite{PhysRevD.108.062001}. This will be particularly important for future high-power interferometers, where alignment imperfections, quadratic mode mismatch, and higher-order wavefront aberrations are expected to become increasingly relevant optical loss mechanisms, especially in configurations employing squeezed-vacuum injection. In particular, alignment and mode-matching sensing schemes, together with the corresponding sensing signal strengths, will be investigated to identify and actively suppress the imperfect coupling degrees of freedom quantified in this work. Such sensing and control strategies will be essential for enabling the practical implementation of higher-order spatial modes in high-precision interferometric experiments, such as future gravitational-wave detectors.

\begin{acknowledgments}
The authors thank Yuefan Guo and Eleonora Capocasa for helpful comments during the preparation of this manuscript. The authors acknowledge support from ANR-18-IDEX-0001 and ANR-23-CE31-0004. This document was submitted to the Virgo and LIGO collaborations under the document numbers VIR-0532A-26 and P2600355, respectively.

\end{acknowledgments}

\appendix

\section{Derivation of the power-loss factors \label{sec-derivation}}

In this appendix, we derive the analytical coupling-loss factors for Laguerre-Gaussian (LG) modes under four small perturbations: angular tilt, lateral displacement, waist size mismatch, and waist position mismatch. For each case, we expand the perturbed field to first order in the corresponding dimensionless imperfection parameter, decompose the perturbation into neighboring LG modes, and sum the powers in the resulting scattered modes.

At the beam waist, $R_c=\infty$, $\Psi=0$, and $w(z)=w_0$. The normalized LG mode is therefore
\begin{equation}
\begin{aligned}
\psi_{p,\ell}(r,\phi;w_0)
={}&
\frac{1}{w_0}
\sqrt{\frac{2p!}{\pi(p+|\ell|)!}}
\left(\frac{\sqrt{2}r}{w_0}\right)^{|\ell|}
\\
&\times
L_p^{|\ell|}\!\left(\frac{2r^2}{w_0^2}\right)
\exp\!\left(-\frac{r^2}{w_0^2}\right)
e^{i\ell\phi}.
\end{aligned}
\end{equation}
The modes satisfy the orthonormality relation
\begin{equation}
\int d^2r\,
\psi_{p',\ell'}^*(r,\phi;w_0)
\psi_{p,\ell}(r,\phi;w_0)
=
\delta_{p'p}\delta_{\ell'\ell}.
\end{equation}

Define the dimensionless radial coordinate
\begin{equation}
u
=
\frac{2r^2}{w_0^2}.
\end{equation}
The LG mode can then be written compactly as
\begin{equation}
\psi_{p,\ell}
=
C_{p,\ell}
w_0^{-(|\ell|+1)}
r^{|\ell|}
L_p^{|\ell|}(u)
e^{-u/2}
e^{i\ell\phi},
\end{equation}
where the factors independent of $w_0$ and $u$ are collected in
\begin{equation}
C_{p,\ell}
=
\sqrt{\frac{2p!}{\pi(p+|\ell|)!}}
\left(\sqrt{2}\right)^{|\ell|}.
\end{equation}

We use the following recurrence and derivative identities for generalized Laguerre polynomials~\cite{AbramowitzStegun1964}:
\begin{equation}
L_p^{|\ell|}(u)
=
L_p^{|\ell|+1}(u)
-
L_{p-1}^{|\ell|+1}(u),
\label{eq-lag-plus}
\end{equation}
\begin{equation}
uL_p^{|\ell|}(u)
=
(p+|\ell|)L_p^{|\ell|-1}(u)
-
(p+1)L_{p+1}^{|\ell|-1}(u),
\label{eq-lag-minus}
\end{equation}
\begin{equation}
\frac{dL_p^{|\ell|}(u)}{du}
=
-L_{p-1}^{|\ell|+1}(u),
\label{eq-lag-deriv}
\end{equation}
and
\begin{equation}
\begin{aligned}
uL_p^{|\ell|}(u)
={}&
(2p+|\ell|+1)L_p^{|\ell|}(u)
-
(p+1)L_{p+1}^{|\ell|}(u)
\\
&-
(p+|\ell|)L_{p-1}^{|\ell|}(u).
\end{aligned}
\label{eq-lag-same}
\end{equation}

We now apply these identities to obtain the scattering amplitudes and coupling losses for each perturbation.

\section*{Misalignment: Tilt}

A small angular tilt $\alpha$ along the Cartesian $x$ direction introduces a linear transverse phase. Define
\begin{equation}
X
=
\frac{x}{w_0}
=
\frac{r\cos\phi}{w_0}.
\end{equation}
The perturbed field is then
\begin{equation}
\psi'_{p,\ell}
=
\psi_{p,\ell}
\exp\left(i k\alpha w_0 X\right),
\end{equation}
where
\begin{equation}
k=\frac{2\pi}{\lambda}.
\end{equation}
In terms of the far-field divergence angle
\begin{equation}
\Theta
=
\frac{\lambda}{\pi w_0},
\end{equation}
the dimensionless tilt parameter is
\begin{equation}
\epsilon_\alpha
=
\frac{\alpha}{\Theta}
=
\frac{k\alpha w_0}{2}.
\label{eq-eps-alpha}
\end{equation}
Hence, the tilted field is
\begin{equation}
\psi'_{p,\ell}
=
\psi_{p,\ell}
\exp\left(i\,2\epsilon_\alpha X\right).
\label{eq-tilt-field}
\end{equation}

Expanding Eq.~\eqref{eq-tilt-field} to first order in $\epsilon_\alpha$ yields
\begin{equation}
\psi'_{p,\ell}
\simeq
\psi_{p,\ell}
+
i\,2\epsilon_\alpha X\psi_{p,\ell}
+
O(\epsilon_\alpha^2).
\label{eq-tilt-exp}
\end{equation}
The first-order term gives the leading amplitudes scattered into orthogonal LG modes.

To decompose $X\psi_{p,\ell}$ in the LG basis, write
\begin{equation}
\psi_{p,\ell}
=
F_{p,\ell}(r)e^{i\ell\phi},
\end{equation}
where
\begin{equation}
F_{p,\ell}(r)
=
C_{p,\ell}
w_0^{-(|\ell|+1)}
r^{|\ell|}
L_p^{|\ell|}(u)
e^{-u/2},
\end{equation}
denotes the radial dependence of the mode. Using
\begin{equation}
x
=
r\cos\phi
=
\frac{r}{2}
\left(
e^{i\phi}+e^{-i\phi}
\right),
\end{equation}
gives
\begin{equation}
X\psi_{p,\ell}
=
\frac{rF_{p,\ell}(r)}{2w_0}
\left[
e^{i(\ell+1)\phi}
+
e^{i(\ell-1)\phi}
\right].
\end{equation}
Thus, multiplication by $X$ changes the azimuthal index $\ell$ by one. We first take $\ell>0$, for which $|\ell|=\ell$ and the radial dependence  can be written directly as $r^\ell L_p^\ell(u)$. The cases $\ell \leq0$ are treated below.

For the term proportional to $e^{i(\ell+1)\phi}$, reducing the radial factor with Eq.~\eqref{eq-lag-plus} and accounting for the LG normalization constants gives
\begin{equation}
\begin{aligned}
&\frac{rF_{p,\ell}(r)}{2w_0}
e^{i(\ell+1)\phi}
\\
&=
\frac{1}{2} \left(\frac{C_{p, \ell}}{C_{p, \ell+1}}\psi_{p, \ell+1} - \frac{C_{p, \ell}}{C_{p-1, \ell+1}}\psi_{p-1, \ell+1} \right) \\
&=
\frac{1}{2\sqrt{2}}
\left[
\sqrt{p+\ell+1}\,
\psi_{p,\ell+1}
-
\sqrt{p}\,
\psi_{p-1,\ell+1}
\right].
\end{aligned}
\label{eq-x-plus}
\end{equation}

For the term proportional to $e^{i(\ell-1)\phi}$, write the additional factor of $r^2$ in terms of $u=2r^2/w_0^2$ and apply Eq.~\eqref{eq-lag-minus}. Including the normalization constants gives
\begin{equation}
\begin{aligned}
&\frac{rF_{p,\ell}(r)}{2w_0}
e^{i(\ell-1)\phi}
\\
&=
\frac{1}{4} \left((p+\ell)\frac{C_{p, \ell}}{C_{p, \ell-1}}\psi_{p, \ell-1} - (p+1)\frac{C_{p, \ell}}{C_{p+1, \ell-1}}\psi_{p+1, \ell-1} \right) \\
&=
\frac{1}{2\sqrt{2}}
\left[
\sqrt{p+\ell}\,
\psi_{p,\ell-1}
-
\sqrt{p+1}\,
\psi_{p+1,\ell-1}
\right].
\end{aligned}
\label{eq-x-minus}
\end{equation}

Combining Eqs.~\eqref{eq-x-plus} and~\eqref{eq-x-minus} gives
\begin{equation}
\begin{aligned}
X\psi_{p,\ell}
=
\frac{1}{2\sqrt{2}}
\Big[
&
\sqrt{p+\ell+1}\,
\psi_{p,\ell+1}
-
\sqrt{p}\,
\psi_{p-1,\ell+1}
\\
&+
\sqrt{p+\ell}\,
\psi_{p,\ell-1}
-
\sqrt{p+1}\,
\psi_{p+1,\ell-1}
\Big],
\end{aligned}
\label{eq-x-decomp}
\end{equation}
where a mode with a negative radial index is understood to be absent.

Substitution of Eq.~\eqref{eq-x-decomp} into Eq.~\eqref{eq-tilt-exp} yields
\begin{equation}
\begin{aligned}
\psi'_{p,\ell}
\simeq{}&
\psi_{p,\ell}
+
\frac{i\epsilon_\alpha}{\sqrt{2}}
\Big[
\sqrt{p+\ell+1}\,
\psi_{p,\ell+1}
-
\sqrt{p}\,
\psi_{p-1,\ell+1}
\\
&\hspace{15mm}
+
\sqrt{p+\ell}\,
\psi_{p,\ell-1}
-
\sqrt{p+1}\,
\psi_{p+1,\ell-1}
\Big]
\\
&+
O(\epsilon_\alpha^2).
\end{aligned}
\end{equation}

Thus, an $x$-directed tilt couples the incident field to four neighboring LG modes of order $\mathcal{N}^{\prime}=\mathcal{N}\pm1$. Their first-order amplitudes are
\begin{equation}
\begin{array}{rlrl}
c_{p,\ell+1}^{(\alpha)}
&=
\dfrac{i\epsilon_\alpha}{\sqrt{2}}
\sqrt{p+\ell+1},
&
c_{p-1,\ell+1}^{(\alpha)}
&=
-\dfrac{i\epsilon_\alpha}{\sqrt{2}}
\sqrt{p},
\\[6pt]
c_{p,\ell-1}^{(\alpha)}
&=
\dfrac{i\epsilon_\alpha}{\sqrt{2}}
\sqrt{p+\ell},
&
c_{p+1,\ell-1}^{(\alpha)}
&=
-\dfrac{i\epsilon_\alpha}{\sqrt{2}}
\sqrt{p+1}.
\end{array}
\label{eq-LG-expansion-alpha}
\end{equation}

For $\ell<0$, the scattered mode decomposition follows from the conjugation symmetry $\psi_{p,-|\ell|}=\psi_{p,|\ell|}^{*}$. Because $X$ is a real operator, the negative-$\ell$ result is obtained by reversing the signs of all azimuthal indices in the corresponding $\ell>0$ result. Denoting the magnitude of the incident azimuthal index by $m=|\ell|$, this correspondence can be written schematically as
\begin{equation}
c_{p',-m\pm1}^{(\alpha)}
=
c_{p',m\mp1}^{(\alpha)},
\qquad m=|\ell|>0.
\label{eq-tilt_negative_l_correspondence}
\end{equation}
where the radial index $p'$ remains the same as in its original branch. Thus, changing the sign of the incident azimuthal index interchanges the $\ell+1$ and $\ell-1$ branches without changing the corresponding complex scattering amplitudes.

The total power scattered out of the incident mode is the sum of the powers in the scattered modes. Because the four scattered modes are mutually orthogonal, their powers add:
\begin{equation}
\begin{aligned}
\mathcal{L}_{\alpha}
&\simeq
P_{p,\ell+1}
+
P_{p-1,\ell+1}
+
P_{p,\ell-1}
+
P_{p+1,\ell-1}
\\
&=
\left|c_{p,\ell+1}^{(\alpha)}\right|^2
+
\left|c_{p-1,\ell+1}^{(\alpha)}\right|^2
+
\left|c_{p,\ell-1}^{(\alpha)}\right|^2
+
\left|c_{p+1,\ell-1}^{(\alpha)}\right|^2
\\
&=
\frac{\epsilon_\alpha^2}{2}
\left[
(p+|\ell|+1)
+p
+(p+|\ell|)
+(p+1)
\right]
\\
&=
\epsilon_\alpha^2
\left(
2p+|\ell|+1
\right).
\end{aligned}
\label{eq-tilt-loss-pos}
\end{equation}

For $\ell=0$, an $x$-directed perturbation produces the symmetric pair of azimuthal indices $\ell'=\pm1$. The radial index of each scattered mode follows from
\begin{equation}
p'
=
\frac{\mathcal{N}'-|\ell'|}{2},
\end{equation}
where $\mathcal{N}'$ and $\ell'$ denote the order and azimuthal index of the scattered mode. The branches $\mathcal{N}'=\mathcal{N}+1$ and $\mathcal{N}'=\mathcal{N}-1$ therefore give $\LGmode{p}{-1}$ and $\LGmode{p-1}{-1}$ for $\ell^\prime=-1$, and $\LGmode{p}{1}$ and $\LGmode{p-1}{1}$ for $\ell^\prime=+1$, respectively. Their relative phases are fixed by the $x$-directed angular dependence $e^{i\phi}+e^{-i\phi}=2\cos\phi$, so the $\ell'=\pm1$ modes within each branch have the same amplitudes. Specifically, at the waist, write
\begin{equation}
    \psi_{p,0}(r)
    =
    C_{p,0} w_0^{-1}L_p^0(u)e^{-u/2},
    \qquad
    u=\frac{2r^2}{w_0^2}.
\end{equation}

The angular-tilt perturbation is proportional to
\(X\psi_{p,0}\). Using
\begin{equation}
\begin{array}{rlrl}
X
&=
\dfrac{r}{2w_0}
\left(
e^{i\phi}+e^{-i\phi}
\right),
&
L_p^0(u)
&=
L_p^1(u)-L_{p-1}^1(u),
\end{array}
\end{equation}

the perturbed field becomes
\begin{equation}
\begin{aligned}
\psi'_{p,0}
\simeq{}&
\psi_{p,0}
+
\frac{i\epsilon_\alpha}{\sqrt{2}}
\Big[
\sqrt{p+1}
\left(
\psi_{p,1}+\psi_{p,-1}
\right)
\\
&\hspace{10mm}
-
\sqrt{p}
\left(
\psi_{p-1,1}+\psi_{p-1,-1}
\right)
\Big]
+
O(\epsilon_\alpha^2),
\end{aligned}
\label{eq-tilt-zero}
\end{equation}

The corresponding scattering coefficients are
\begin{equation}
\begin{array}{rlrl}
c_{p,\pm1}^{(\alpha)}
&=
\dfrac{i\epsilon_\alpha}{\sqrt{2}}
\sqrt{p+1},
&
c_{p-1,\pm1}^{(\alpha)}
&=
-\dfrac{i\epsilon_\alpha}{\sqrt{2}}
\sqrt{p},
\end{array}
\end{equation}
Here, $c_{p,\pm1}$ denotes equal coefficients for the $\ell'=+1$ and $\ell'=-1$ modes; any mode with a negative radial index is absent.

Summing the powers in both helicity branches gives
\begin{equation}
\begin{aligned}
\mathcal{L}_{\alpha} |_{\ell=0}
&\simeq
2\left[
\frac{\epsilon_{\alpha}^2}{2}(p+1)
\right]
+
2\left[
\frac{\epsilon_{\alpha}^2}{2}p
\right]
=
\epsilon_{\alpha}^2
\left(
2p+1
\right).
\end{aligned}
\end{equation}
This result agrees with Eq.~\eqref{eq-tilt-loss-pos} at $\ell=0$.

For the fundamental Gaussian mode, $p=\ell=0$, the terms with radial index $p-1$ are absent, and Eq.~\eqref{eq-tilt-zero} reduces to
\begin{equation}
\begin{array}{rl}
\psi'_{0,0}
&\simeq
\psi_{0,0}
+
\dfrac{i\epsilon_\alpha}{\sqrt{2}}
\left(
\psi_{0,1}+\psi_{0,-1}
\right)
+
O(\epsilon_\alpha^2).
\end{array}
\end{equation}
Since $\HGmode{1}{0}$ mode can be expressed in the LG basis as
\begin{equation}
\psi_{10}^{\mathrm{HG}}
=
\frac{1}{\sqrt{2}}
\left(
\psi_{0,1}+\psi_{0,-1}
\right),
\end{equation}
these expressions become
\begin{equation}
\begin{array}{rlrl}
\psi'_{0,0}
&\simeq
\psi_{0,0}
+
i\epsilon_\alpha\psi_{10}^{\mathrm{HG}},
&
\mathcal{L}_{\alpha} |_{(0,0)}
&\simeq
\epsilon_\alpha^2.
\end{array}
\end{equation}
This is the standard first-order result for a tilted fundamental Gaussian beam~\cite{Anderson:84}.

The total loss depends only on $|\ell|$ and can therefore be written for arbitrary $\ell$ is
\begin{equation}
\mathcal{L}_{\alpha}
\simeq
\epsilon_\alpha^2
\left(
2p+|\ell|+1
\right).
\label{eq-tilt-all-ell}
\end{equation}

Using Eq.~\eqref{eq-eps-alpha}, the angular tilt-induced power loss can equivalently be expressed as
\begin{equation}
\begin{aligned}
\mathcal{L}_{\alpha}
&\simeq
\frac{k^2\alpha^2w_0^2}{4}
\left(
2p+|\ell|+1
\right)
\\
&=
\left(
\frac{\pi w_0\alpha}{\lambda}
\right)^2
\left(
2p+|\ell|+1
\right).
\end{aligned}
\end{equation}

\section*{Misalignment: Lateral offset}

For a lateral offset $\Delta x$ along the Cartesian $x$ direction, the perturbed field is
\begin{equation}
\psi'_{p,\ell}(x,y)
=
\psi_{p,\ell}(x-\Delta x,y).
\label{eq-offset-field}
\end{equation}
Define the dimensionless offset parameter
\begin{equation}
\epsilon_a
=
\frac{\Delta x}{w_0}.
\label{eq-eps-a}
\end{equation}
Expanding Eq.~\eqref{eq-offset-field} to first order in $\Delta x$ gives
\begin{equation}
\psi'_{p,\ell}
\simeq
\psi_{p,\ell}
-
\epsilon_a w_0
\frac{\partial\psi_{p,\ell}}{\partial x}
+
O(\epsilon_a^2).
\label{eq-offset-exp}
\end{equation}

To decompose $\partial_x\psi_{p,\ell}$ in the LG basis, first take $\ell>0$ and write
\begin{equation}
\psi_{p,\ell}
=
F_{p,\ell}(r)e^{i\ell\phi},
\end{equation}
where $F_{p,\ell}(r)$ is the radial part. The Cartesian derivative is
\begin{equation}
\frac{\partial}{\partial x}
=
\cos\phi\frac{\partial}{\partial r}
-
\frac{\sin\phi}{r}
\frac{\partial}{\partial\phi}.
\end{equation}
Therefore,
\begin{equation}
\begin{aligned}
\frac{\partial\psi_{p,\ell}}{\partial x}
={}&
\cos\phi\,
\frac{dF_{p,\ell}}{dr}
e^{i\ell\phi}
-
\frac{i\ell\sin\phi}{r}
F_{p,\ell}(r)e^{i\ell\phi}.
\end{aligned}
\end{equation}
Using
\begin{equation}
\begin{array}{rlrl}
\cos\phi
&=
\dfrac{1}{2}
\left(
e^{i\phi}+e^{-i\phi}
\right),
&
\sin\phi
&=
\dfrac{1}{2i}
\left(
e^{i\phi}-e^{-i\phi}
\right).
\end{array}
\end{equation}
It follows that
\begin{equation}
\begin{aligned}
\frac{\partial\psi_{p,\ell}}{\partial x}
={}&
\frac{1}{2}
\left(
\frac{dF_{p,\ell}}{dr}
-
\frac{\ell}{r}F_{p,\ell}
\right)
e^{i(\ell+1)\phi}
\\
&+
\frac{1}{2}
\left(
\frac{dF_{p,\ell}}{dr}
+
\frac{\ell}{r}F_{p,\ell}
\right)
e^{i(\ell-1)\phi}.
\end{aligned}
\label{eq-dx-angular}
\end{equation}
Similar to the angular tilt case, differentiation with respect to $x$ changes the azimuthal index by one.

Differentiating the radial function yields
\begin{equation}
\begin{aligned}
\frac{dF_{p,\ell}}{dr}
&={}
C_{p,\ell}w_0^{-(\ell+1)}
r^{\ell-1}e^{-u/2} 
\\ 
&\times
\Bigg[
\ell L_p^\ell(u)
+
2u\frac{dL_p^\ell(u)}{du}
-
uL_p^\ell(u)
\Bigg].
\end{aligned}
\label{eq-radial-deriv}
\end{equation}

For the term proportional to $e^{i(\ell+1)\phi}$, the derivative of $r^\ell$ in Eq.~\eqref{eq-radial-deriv} cancels the $\ell/r$ term in Eq.~\eqref{eq-dx-angular}. Reducing the remainder with Eqs.~\eqref{eq-lag-deriv} and~\eqref{eq-lag-plus}, and including the normalization constants, gives
\begin{equation}
\begin{aligned}
\frac{1}{2}
&\left(
\frac{dF_{p,\ell}}{dr}
-
\frac{\ell}{r}F_{p,\ell}
\right)
e^{i(\ell+1)\phi}
\\
&=-\frac{1}{w_0} \Big[\frac{C_{p, \ell}}{C_{p-1, \ell+1}} \psi_{p-1,\ell+1} + \frac{C_{p, \ell}}{C_{p, \ell+1}} \psi_{p,\ell+1} \Big]
\\
&=
-\frac{1}{\sqrt{2}w_0}
\Big[
\sqrt{p}\,
\psi_{p-1,\ell+1}
+
\sqrt{p+\ell+1}\,
\psi_{p,\ell+1}
\Big].
\end{aligned}
\label{eq-dx-plus}
\end{equation}

Similarly, for the term proportional to $e^{i(\ell-1)\phi}$, using Eqs.~\eqref{eq-lag-deriv} and~\eqref{eq-lag-minus} gives
\begin{equation}
\begin{aligned}
\frac{1}{2}
&\left(
\frac{dF_{p,\ell}}{dr}
+
\frac{\ell}{r}F_{p,\ell}
\right)
e^{i(\ell-1)\phi}
\\
&= \frac{1}{2 w_0} \Big[(p+\ell)\frac{C_{p, \ell}}{C_{p, \ell-1}} \psi_{p, \ell-1}  + (p+1)\frac{C_{p, \ell}}{C_{p+1, \ell-1}}\psi_{p+1, \ell-1} \Big] 
\\
&=
\frac{1}{\sqrt{2}w_0}
\Big[
\sqrt{p+\ell}\,
\psi_{p,\ell-1}
+
\sqrt{p+1}\,
\psi_{p+1,\ell-1}
\Big].
\end{aligned}
\label{eq-dx-minus}
\end{equation}

Combining Eqs.~\eqref{eq-dx-plus} and~\eqref{eq-dx-minus} gives
\begin{equation}
\begin{aligned}
\frac{\partial\psi_{p,\ell}}{\partial x}
=
\frac{1}{\sqrt{2}w_0}
\Big[
&
-\sqrt{p}\,
\psi_{p-1,\ell+1}
-
\sqrt{p+\ell+1}\,
\psi_{p,\ell+1}
\\
&+
\sqrt{p+\ell}\,
\psi_{p,\ell-1}
+
\sqrt{p+1}\,
\psi_{p+1,\ell-1}
\Big],
\end{aligned}
\label{eq-dx-decomp}
\end{equation}
where a mode with a negative radial index is understood to be absent.

Substitution of Eq.~\eqref{eq-dx-decomp} into Eq.~\eqref{eq-offset-exp} yields
\begin{equation}
\begin{aligned}
\psi'_{p,\ell}
&\simeq{}
\psi_{p,\ell}
+
\frac{\epsilon_a}{\sqrt{2}}
\Big[
\sqrt{p+\ell+1}\,
\psi_{p,\ell+1}
+
\sqrt{p}\,
\psi_{p-1,\ell+1}
\\
&
-
\sqrt{p+\ell}\,
\psi_{p,\ell-1}
-
\sqrt{p+1}\,
\psi_{p+1,\ell-1}
\Big]
+
O(\epsilon_a^2).
\end{aligned}
\end{equation}

Thus, similar to the angular tilt case, an $x$-directed lateral offset couples the incident field to four neighboring LG modes of order $\mathcal{N}^{\prime}=\mathcal{N}\pm1$. Their first-order amplitudes are
\begin{equation}
\begin{array}{rlrl}
c_{p,\ell+1}^{(a)}
&=
\dfrac{\epsilon_a}{\sqrt{2}}
\sqrt{p+\ell+1},
&
c_{p-1,\ell+1}^{(a)}
&=
\dfrac{\epsilon_a}{\sqrt{2}}
\sqrt{p},
\\[6pt]
c_{p,\ell-1}^{(a)}
&=
-\dfrac{\epsilon_a}{\sqrt{2}}
\sqrt{p+\ell},
&
c_{p+1,\ell-1}^{(a)}
&=
-\dfrac{\epsilon_a}{\sqrt{2}}
\sqrt{p+1}.
\end{array}
\label{eq-LG-expansion-a}
\end{equation}

For $\ell<0$, the mode decomposition follows from the same conjugation symmetry argument as in the tilt case. Accordingly, the correspondence in Eq.~\eqref{eq-tilt_negative_l_correspondence} also holds for lateral translation:
\begin{equation}
c_{p',-m\pm1}^{(a)}
=
c_{p',m\mp1}^{(a)},
\qquad m=|\ell|>0.
\end{equation}
The coefficient amplitudes are unchanged.

The scattered-mode powers add to
\begin{equation}
\begin{aligned}
\mathcal{L}_{a}
&\simeq
P_{p,\ell+1}
+
P_{p-1,\ell+1}
+
P_{p,\ell-1}
+
P_{p+1,\ell-1}
\\
&=
\left|c_{p,\ell+1}^{(a)}\right|^2
+
\left|c_{p-1,\ell+1}^{(a)}\right|^2
+
\left|c_{p,\ell-1}^{(a)}\right|^2
+
\left|c_{p+1,\ell-1}^{(a)}\right|^2
\\
&=
\frac{\epsilon_a^2}{2}
\left[
(p+|\ell|+1)
+p
+(p+|\ell|)
+(p+1)
\right]
\\
&=
\epsilon_a^2
\left(
2p+|\ell|+1
\right).
\end{aligned}
\end{equation}

For $\ell=0$, an $x$-directed offset produces the symmetric pair $\ell'=\pm1$. Using
\begin{equation}
\begin{array}{rlrl}
\dfrac{dL_p^0(u)}{du}
&=
-L_{p-1}^1(u),
&
L_p^0(u)
&=
L_p^1(u)-L_{p-1}^1(u),
\end{array}
\end{equation}
gives
\begin{equation}
\begin{aligned}
\psi'_{p,0}
\simeq{}&
\psi_{p,0}
+
\frac{\epsilon_a}{\sqrt{2}}
\Big[
\sqrt{p+1}
\left(
\psi_{p,1}+\psi_{p,-1}
\right)
\\
&\hspace{10mm}
+
\sqrt{p}
\left(
\psi_{p-1,1}+\psi_{p-1,-1}
\right)
\Big]
+
O(\epsilon_a^2).
\end{aligned}
\label{eq-offset-zero}
\end{equation}

The corresponding scattering coefficients are
\begin{equation}
\begin{array}{rlrl}
c_{p,\pm1}^{(a)}
&=
\dfrac{\epsilon_a}{\sqrt{2}}
\sqrt{p+1},
&
c_{p-1,\pm1}^{(a)}
&=
\dfrac{\epsilon_a}{\sqrt{2}}
\sqrt{p},
\end{array}
\end{equation}

Summing the powers in both helicity branches gives
\begin{equation}
\begin{aligned}
\mathcal{L}_{a}
&\simeq
2\left[
\frac{\epsilon_{a}^2}{2}(p+1)
\right]
+
2\left[
\frac{\epsilon_{a}^2}{2}p
\right]
=
\epsilon_{a}^2
\left(
2p+1
\right).
\end{aligned}
\end{equation}
This agrees with the general loss expression at $\ell=0$.

For the fundamental Gaussian mode, $p=\ell=0$, the terms with radial index $p-1$ are absent, and Eq.~\eqref{eq-offset-zero} reduces to
\begin{equation}
\begin{array}{rl}
\psi'_{0,0}
&\simeq
\psi_{0,0}
+
\dfrac{\epsilon_a}{\sqrt{2}}
\left(
\psi_{0,1}+\psi_{0,-1}
\right)
+
O(\epsilon_a^2),
\end{array}
\end{equation}
These expressions become
\begin{equation}
\begin{array}{rlrl}
\psi'_{0,0}
&\simeq
\psi_{0,0}
+
\epsilon_a\psi_{10}^{\mathrm{HG}},
&
\mathcal{L}_{a} |_{(0,0)}
&\simeq
\epsilon_a^2.
\end{array}
\end{equation}
This is the standard first-order result for a laterally displaced fundamental Gaussian beam~\cite{Anderson:84}.

Since the loss depends only on $|\ell|$, the result for arbitrary $\ell$ is
\begin{equation}
\mathcal{L}_{a}
\simeq
\epsilon_a^2
\left(
2p+|\ell|+1
\right).
\label{eq-lateral-all-ell}
\end{equation}

Using Eq.~\eqref{eq-eps-a}, the loss can also be written as
\begin{equation}
\mathcal{L}_{a}
\simeq
\left(
\frac{\Delta x}{w_0}
\right)^2
\left(
2p+|\ell|+1
\right).
\end{equation}

\section*{Mode mismatch: Waist size}

Consider a small mismatch between the reference waist size $w_0$ and the perturbed waist size $w_0'$, written as
\begin{equation}
w_0'
=
w_0(1+\epsilon_w),
\qquad
\epsilon_w
=
\frac{w_0'}{w_0}-1.
\end{equation}
The waist-size change is therefore
\begin{equation}
\delta w
=
w_0'-w_0
=
\epsilon_w w_0.
\end{equation}
Expanding about the reference waist size gives
\begin{equation}
\psi_{p,\ell}(w_0')
\simeq
\psi_{p,\ell}(w_0)
+
\epsilon_w w_0
\frac{\partial\psi_{p,\ell}}{\partial w_0}
+
O(\epsilon_w^2).
\label{eq-waist-exp}
\end{equation}

To decompose $\partial\psi_{p,\ell}/\partial w_0$ in the LG basis, note that $u=\frac{2r^2}{w_0^2}$ satisfies
\begin{equation}
w_0\frac{\partial u}{\partial w_0}
=
-2u.
\end{equation}
Using the compact waist-plane form
\begin{equation}
\psi_{p,\ell}
=
C_{p,\ell}
w_0^{-(|\ell|+1)}
r^{|\ell|}
L_p^{|\ell|}(u)
e^{-u/2}
e^{i\ell\phi},
\end{equation}
gives
\begin{equation}
\begin{aligned}
w_0\frac{\partial\psi_{p,\ell}}{\partial w_0}
={}&
C_{p,\ell}r^{|\ell|}e^{i\ell\phi}
w_0\frac{\partial}{\partial w_0}
\left[
w_0^{-(|\ell|+1)}
L_p^{|\ell|}(u)e^{-u/2}
\right].
\end{aligned}
\end{equation}

The three $w_0$-dependent factors satisfy
\begin{equation}
w_0
\frac{\partial}{\partial w_0}
w_0^{-(|\ell|+1)}
=
-(|\ell|+1)
w_0^{-(|\ell|+1)},
\end{equation}
\begin{equation}
\begin{aligned}
w_0
\frac{\partial L_p^{|\ell|}(u)}{\partial w_0}
&=
-2u
\frac{dL_p^{|\ell|}(u)}{du}
=
2uL_{p-1}^{|\ell|+1}(u),
\end{aligned}
\end{equation}
where Eq.~\eqref{eq-lag-deriv} was used, and
\begin{equation}
w_0
\frac{\partial e^{-u/2}}{\partial w_0}
=
u e^{-u/2}.
\end{equation}
Combining the three contributions gives
\begin{equation}
\begin{aligned}
w_0\frac{\partial\psi_{p,\ell}}{\partial w_0}
={}&
C_{p,\ell}
w_0^{-(|\ell|+1)}
r^{|\ell|}
e^{-u/2}
e^{i\ell\phi}
\\
&\times
\left[
(u-|\ell|-1)L_p^{|\ell|}(u)
+
2uL_{p-1}^{|\ell|+1}(u)
\right].
\end{aligned}
\label{eq-dw-pre}
\end{equation}

Applying Eqs.~\eqref{eq-lag-same} and~\eqref{eq-lag-minus} to the radial factor in Eq.~\eqref{eq-dw-pre} gives
\begin{equation}
\begin{aligned}
&(u-|\ell|-1)L_p^{|\ell|}(u)
+
2uL_{p-1}^{|\ell|+1}(u)
\\
={}&
\left[
(2p+|\ell|+1)L_p^{|\ell|}(u)
-
(p+1)L_{p+1}^{|\ell|}(u)
\right.
\\
&\left.\hspace{0mm}
-
(p+|\ell|)L_{p-1}^{|\ell|}(u)
\right]
-
(|\ell|+1)L_p^{|\ell|}(u)
\\
&+
2\left[
(p+|\ell|)L_{p-1}^{|\ell|}(u)
-
pL_p^{|\ell|}(u)
\right]
\\
={}&
-(p+1)L_{p+1}^{|\ell|}(u)
+
(p+|\ell|)L_{p-1}^{|\ell|}(u).
\end{aligned}
\end{equation}
The contribution proportional to the original radial polynomial $L_p^{|\ell|}(u)$ cancels exactly, leaving
\begin{equation}
\begin{aligned}
&w_0\frac{\partial\psi_{p,\ell}}{\partial w_0}
\\
&= -(p+1)\frac{C_{p, \ell}}{C_{p+1, \ell}} \psi_{p+1, \ell} + (p+|\ell|)\frac{C_{p, \ell}}{C_{p-1, \ell}} \psi_{p-1, \ell}
\\
&
= -\sqrt{(p+1)(p+|\ell|+1)}\,
\psi_{p+1,\ell}
+
\sqrt{p(p+|\ell|)}\,
\psi_{p-1,\ell}.
\end{aligned}
\label{eq-dw-decomp}
\end{equation}
For $p=0$, the mode $\psi_{p-1,\ell}$ is absent. Equation~\eqref{eq-dw-decomp} shows that a cylindrically symmetric waist size mismatch preserves the azimuthal index and couples only to radial indices $p\pm1$, whose mode orders are $\mathcal{N}^\prime=\mathcal{N}\pm2$.

Substitution of Eq.~\eqref{eq-dw-decomp} into Eq.~\eqref{eq-waist-exp} yields
\begin{equation}
\begin{aligned}
\psi_{p,\ell}(w_0')
\simeq{}&
\psi_{p,\ell}(w_0)
-
\epsilon_w
\sqrt{(p+1)(p+|\ell|+1)}\,
\psi_{p+1,\ell}
\\
&+
\epsilon_w
\sqrt{p(p+|\ell|)}\,
\psi_{p-1,\ell}
+
O(\epsilon_w^2).
\end{aligned}
\end{equation}

The first-order scattering amplitudes are
\begin{equation}
\begin{array}{rl}
c_{p+1,\ell}^{(w)}
&=
-\epsilon_w
\sqrt{(p+1)(p+|\ell|+1)},
\\[6pt]
c_{p-1,\ell}^{(w)}
&=
\epsilon_w
\sqrt{p(p+|\ell|)}.
\end{array}
\label{eq-LG-expansion-w}
\end{equation}

The scattered mode powers add to
\begin{equation}
\begin{aligned}
\mathcal{L}_w
&\simeq
P_{p+1,\ell}^{(w)}
+
P_{p-1,\ell}^{(w)}
\\
&=
\left|c_{p+1,\ell}^{(w)}\right|^2
+
\left|c_{p-1,\ell}^{(w)}\right|^2
\\
&=
\epsilon_w^2
\left[
(p+1)(p+|\ell|+1)
+
p(p+|\ell|)
\right]
\\
&=
\epsilon_w^2
\left[
2p^2+2p+(2p+1)|\ell|+1
\right].
\end{aligned}
\label{eq-ws-all-ell}
\end{equation}

For the special donut-shaped $\LGmode{0}{\ell}$ mode family, only $\LGmode{1}{\ell}$ is generated at first order:
\begin{equation}
\psi_{0,\ell}(w_0')
\simeq
\psi_{0,\ell}(w_0)
-
\epsilon_w\sqrt{|\ell|+1}\,
\psi_{1,\ell}
+
O(\epsilon_w^2).
\end{equation}
The corresponding power loss is
\begin{equation}
\mathcal{L}_w |_{p=0}
\simeq
(|\ell|+1)\epsilon_w^2.
\end{equation}

For the fundamental Gaussian mode, $p=\ell=0$, these expressions reduce to
\begin{equation}
\begin{array}{rlrl}
\psi'_{0,0}
&\simeq
\psi_{0,0}
- \epsilon_w \psi_{1,0},
&
\mathcal{L}_{\mathrm{w}} |_{(0,0)}
&\simeq
\epsilon_w^2.
\end{array}
\end{equation}
This is the standard first-order result for waist size mismatch of a fundamental Gaussian beam~\cite{Anderson:84}.

\section*{Mode mismatch: Waist position}

A small mismatch in the waist position produces a residual quadratic phase at the reference waist plane. The perturbed field is
\begin{equation}
\psi'_{p,\ell}
=
\psi_{p,\ell}
\exp\left(
-i\frac{kr^2}{2R_c}
\right),
\end{equation}
where $R_c$ is the wavefront radius of curvature of the mismatched beam at the reference waist plane. For a small longitudinal waist displacement $\delta z_0$,
\begin{equation}
\frac{1}{R_c}
=
\frac{\delta z_0}{\delta z_0^2+z_R^2}
\simeq
\frac{\delta z_0}{z_R^2},
\end{equation}
where $z_R=\frac{\pi w_0^2}{\lambda}$ is the Rayleigh range. Thus,
\begin{equation}
\psi'_{p,\ell}
=
\psi_{p,\ell}
\exp\left(
-i\frac{kr^2}{2z_R^2}\delta z_0
\right) 
= \psi_{p,\ell}
\exp\left(
-iu\epsilon_z
\right),
\label{eq-z-field}
\end{equation}
where $u=\frac{2r^2}{w_0^2}$ and the dimensionless waist position mismatch parameter is
\begin{equation}
\epsilon_z
=
\frac{\delta z_0}{2z_R},
\end{equation}

Expanding Eq.~\eqref{eq-z-field} to first order in $\epsilon_z$ yields
\begin{equation}
\psi'_{p,\ell}
\simeq
\psi_{p,\ell}
-
i\epsilon_z u\psi_{p,\ell}
+
O(\epsilon_z^2).
\label{eq-z-exp}
\end{equation}

Applying Eq.~\eqref{eq-lag-same} decomposes $u\psi_{p,\ell}$ in the LG basis:
\begin{equation}
\begin{aligned}
u\psi_{p,\ell}
={}&
(2p+|\ell|+1)\psi_{p,\ell} -
(p+1)
\frac{C_{p,\ell}}{C_{p+1,\ell}}
\psi_{p+1,\ell}
\\
&-
(p+|\ell|)
\frac{C_{p,\ell}}{C_{p-1,\ell}}
\psi_{p-1,\ell} 
\\
&=
(2p+|\ell|+1)\psi_{p,\ell}
\\
&-
\sqrt{(p+1)(p+|\ell|+1)}\,
\psi_{p+1,\ell}
\\
&-
\sqrt{p(p+|\ell|)}\,
\psi_{p-1,\ell}.
\end{aligned}
\label{eq-u-decomp}
\end{equation}
For $p=0$, the mode $\psi_{p-1,\ell}$ is absent.

Similar to the waist size mismatch, Eq.~\eqref{eq-u-decomp} shows that the quadratic phase perturbation associated with waist position mismatch preserves the azimuthal index $\ell$ and couples at first order only to neighboring radial indices $p\pm1$, whose mode orders are $\mathcal{N}^\prime=\mathcal{N}\pm2$.

Substitution of Eq.~\eqref{eq-u-decomp} into Eq.~\eqref{eq-z-exp} yields
\begin{equation}
\begin{aligned}
\psi'_{p,\ell}
\simeq{}&
\left[
1
-
i\epsilon_z
(2p+|\ell|+1)
\right]
\psi_{p,\ell}
\\
&+
i\epsilon_z
\sqrt{(p+1)(p+|\ell|+1)}\,
\psi_{p+1,\ell}
\\
&+
i\epsilon_z
\sqrt{p(p+|\ell|)}\,
\psi_{p-1,\ell}
+
O(\epsilon_z^2).
\end{aligned}
\end{equation}

The first-order scattering amplitudes are
\begin{equation}
\begin{array}{rl}
c_{p+1,\ell}^{(z)}
&=
i\epsilon_z
\sqrt{(p+1)(p+|\ell|+1)},
\\[6pt]
c_{p-1,\ell}^{(z)}
&=
i\epsilon_z
\sqrt{p(p+|\ell|)}.
\end{array}
\label{eq-LG-expansion-z}
\end{equation}
The first-order change in the amplitude of the original mode is
\begin{equation}
\delta c_{p,\ell}^{(z)}
=
-i\epsilon_z
(2p+|\ell|+1),
\end{equation}
which is purely imaginary and therefore represents a first-order phase shift of the original mode rather than a change in its power. Consequently, it does not contribute to the leading-order power loss. The total power loss is therefore
\begin{equation}
\begin{aligned}
\mathcal{L}_z
&\simeq
P_{p+1,\ell}^{(z)}
+
P_{p-1,\ell}^{(z)}
\\
&=
\left|c_{p+1,\ell}^{(z)}\right|^2
+
\left|c_{p-1,\ell}^{(z)}\right|^2
\\
&=
\epsilon_z^2
\left[
(p+1)(p+|\ell|+1)
+
p(p+|\ell|)
\right]
\\
&=
\epsilon_z^2
\left[
2p^2+2p+(2p+1)|\ell|+1
\right].
\end{aligned}
\label{eq-wp-all-ell}
\end{equation}

For the special donut-shaped $\LGmode{0}{\ell}$ mode family, only $\LGmode{1}{\ell}$ is generated at first order:
\begin{equation}
\begin{aligned}
\psi'_{0,\ell}
\simeq{}&
\left[
1
-
i\epsilon_z(|\ell|+1)
\right]
\psi_{0,\ell}
+
i\epsilon_z
\sqrt{|\ell|+1}\,
\psi_{1,\ell}
+
O(\epsilon_z^2).
\end{aligned}
\end{equation}
The corresponding power loss is
\begin{equation}
\mathcal{L}_z |_{p=0}
\simeq
(|\ell|+1)\epsilon_z^2.
\end{equation}

For the fundamental Gaussian mode, $p=\ell=0$, these expressions reduce to
\begin{equation}
\begin{array}{rlrl}
\psi'_{0,0}
&\simeq
(1-i\epsilon_z) \psi_{0,0}
+ i\epsilon_z \psi_{1,0},
&
\mathcal{L}_{\mathrm{z}}|_{(0,0)}
&\simeq
\epsilon_z^2.
\end{array}
\end{equation}
This is the standard first-order result for waist position mismatch of a fundamental Gaussian beam~\cite{Anderson:84}.

Together, these results complete the analytical derivation of the leading-order scattering into neighboring modes and the corresponding power loss factors for generic $\LGmode{p}{\ell}$ modes under all misalignment and mode mismatch degrees of freedom considered in this work.

\bibliography{references}

\end{document}